\RequirePackage[running]{lineno}

\documentclass[aps,prl,nofootinbib,twocolumn,showpacs,superscriptaddress,letterpaper,amsmath,amssymb]{revtex4}

\usepackage{graphicx}  % needed for figures
\usepackage{dcolumn}   % needed for some tables
\usepackage{bm}        % for math
\usepackage{amssymb}   % for math
\usepackage{feynmf}%%%{feynmp}
\usepackage{slashed}
\usepackage{multirow}
\def\gsim{\lower0.5ex\hbox{$\:\buildrel >\over\sim\:$}}
\def\lsim{\lower0.5ex\hbox{$\:\buildrel <\over\sim\:$}}

\newcommand{\be}{\begin{equation}}
\newcommand{\ee}{\end{equation}}
\newcommand{\bea}{\begin{eqnarray}}
\newcommand{\eea}{\end{eqnarray}}

\newcommand{\nbox}{{\,\lower0.9pt\vbox{\hrule \hbox{\vrule height 0.2 cm
\hskip 0.2 cm \vrule height 0.2 cm}\hrule}\,}}

\def\ie{{\it i.e.}}
\def\eg{{\it e.g.}}

\def\sub#1{_{\lower.25ex\hbox{$\scriptstyle#1$}}}

\def\gev{\,{\rm GeV}}

\def\to{\rightarrow}

\newskip\zatskip \zatskip=0pt plus0pt minus0pt
\def\matth{\mathsurround=0pt}
\def\lsim{\mathrel{\mathpalette\atversim<}}
\def\gsim{\mathrel{\mathpalette\atversim>}}
\def\sigv{\ifmmode \langle\sigma v\rangle\else $\langle\sigma v\rangle$\fi}
\newskip\zatskip \zatskip=0pt plus0pt minus0pt
\def\matth{\mathsurround=0pt}
\def\lsim{\mathrel{\mathpalette\atversim<}}
\def\gsim{\mathrel{\mathpalette\atversim>}}
\def\atversim#1#2{\lower0.7ex\vbox{\baselineskip\zatskip\lineskip\zatskip
  \lineskiplimit
  0pt\ialign{$\matth#1\hfil##\hfil$\crcr#2\crcr\sim\crcr}}}

\def\missET {{\not\!\! E_T}}

\begin{document}

\thispagestyle{empty}
\vspace*{-3.5cm}

\vspace{0.5in}

%\begin{flushright}
%\today\\
%\end{flushright}
%\vspace{0.5in}
\title{Confronting the Fermi Line with LHC data:\\ an Effective Theory
  of Dark Matter Interaction with Photons}

\begin{center}
\begin{abstract}
We describe an effective theory of interaction between pairs of dark
matter particles and pairs of photons. Such an interaction could
accomodate $\chi\bar{\chi}\rightarrow\gamma\gamma$ processes which might be
the cause of the observed feature in the FermiLAT spectrum, as well as
$\gamma^*/Z\rightarrow \gamma\chi\bar{\chi}$ processes, which would
predict excesses at the LHC in the $\gamma+\missET$ final-state.  We reinterpret
an ATLAS $\gamma+\missET$ analysis and the observed Fermi feature in
the parameter space of our new effective theory to assess their consistency.
\end{abstract}
\end{center}

\author{Andy Nelson}
\affiliation{Department of Physics and Astronomy, University of California, Irvine, CA 92697}
\author{Linda M. Carpenter}
\affiliation{The Ohio State University, Columbus, OH}
\author{Randel Cotta}
\affiliation{Department of Physics and Astronomy, University of
  California, Irvine, CA 92697}
\author{Adam Johnstone}
\affiliation{Department of Physics and Astronomy, University of
  California, Irvine, CA 92697}
\author{Daniel Whiteson}
\affiliation{Department of Physics and Astronomy, University of
  California, Irvine, CA 92697}

%\preprint{UCI-HEP-TR-2012-XX}
\maketitle

% introduction
\linenumbers

\subsection{Introduction}

Strong evidence for dark matter exists in the form of precise measurements of 
galactic rotation curves and gravitational lensing, but its  nature is still largely 
a mystery~\cite{dmReview}.   A vigorous experimental program seeks to
identify the particle nature of weakly-interacting cold dark
matter, $\chi$, by looking for the
scattering of heavy nuclei by local dark matter, or annhilation in space of dark matter pairs into standard model particles. Recently,
a statistically significant peak was observed in the Fermi-LAT photon
spectrum, which can be interpretted as $\chi\chi\rightarrow
\gamma\gamma$~\cite{fermiLatExcess}, though concerns have been raised
about its possible origin as an instrumental artifact~\cite{fermisyst1,fermisyst2,fermisyst3}.

\begin{figure}
\includegraphics[width=1.5in]{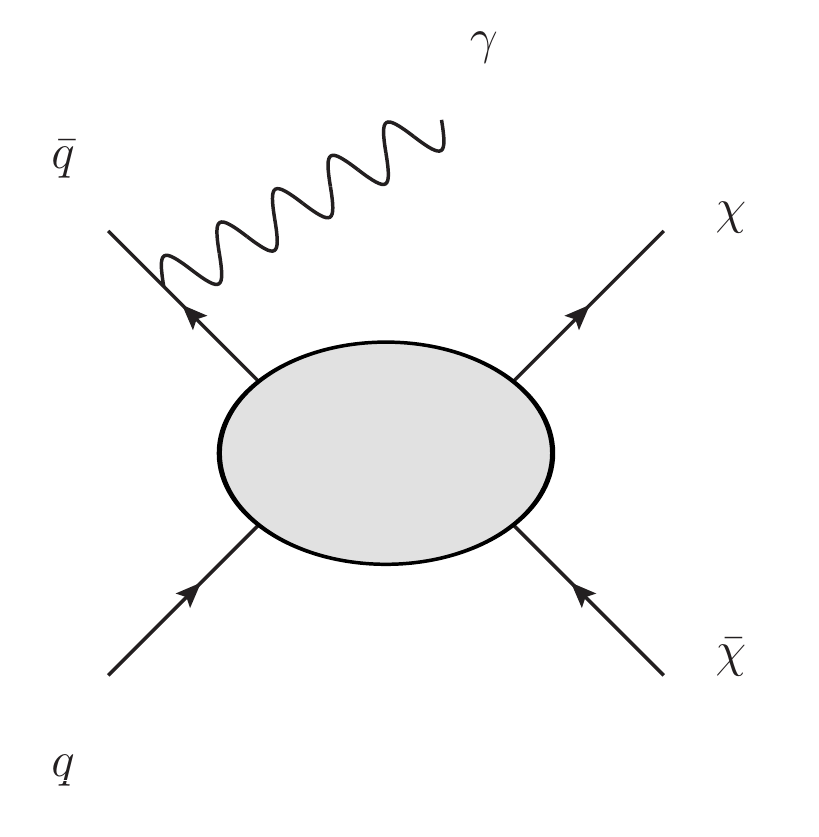}\\
\includegraphics[width=1.65in]{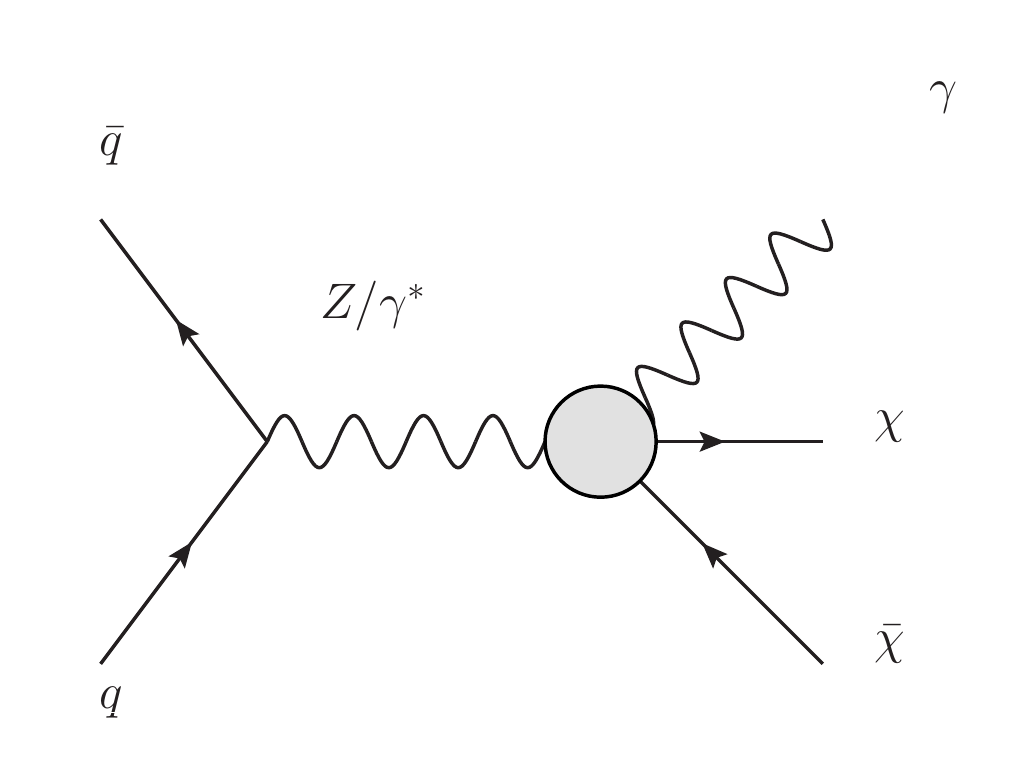}
\caption{\label{diag}Representative diagrams for production of dark matter pairs ($\chi \bar{\chi}$) associated with
a photon in theories where dark matter interacts with quarks (top) or directly with weak boson pairs (bottom). The latter
are those that we consider in this work.}
\end{figure}

 High
energy particle accelerators provide another experimental probe, as
they can directly produce  pairs of dark matter particles, independent
of the local or galactic dark matter density. Pairs of dark matter particles produced at colliders are, however, invisible to
the detectors. A fruitful approach has been to consider the case in
which a standard model particle is emitted as initial state radiation
preceeding the dark matter production, see the top of Fig.~\ref{diag}. The final state signature is
then a single reconstructed object (jet~\cite{atlasjet,cmsjet},
photon~\cite{atlasphoton,cmsphoton}, $Z$ boson~\cite{monoz,Bell:2012rg} etc) with no object to balance its
transverse momentum, leading to large missing transverse momentum
($\missET$).

The production of dark matter particles is usually assumed to be due to an
interaction between the dark matter particles $\chi$ and the primary constituents of the collider
initial state ($q$ or $g$).  The precise nature of this interaction is not
known, but a useful general formalism is provided by effective field
theories~\cite{Goodman:2010ku,Beltran:2010ww,Fox:2011pm}, which are strictly speaking
valid only when the coupling occurs through states which are heavy compared with the typical
energies involved ($\sim\missET$) and can be integrated out to give an effective
four-fermion operator.

Recently~\cite{monoz,Cotta:2012nj}, this has been extended to consider
the case of an effective field theory which couples
the dark matter fields to electroweak bosons rather than to the
fermionic initial state.  For such interactions the $Z$ boson+$\missET$
final state would be one of the unique strategies for searching for dark
matter at colliders.

In this paper, we extend this line of thought to the $\gamma+\missET$
final state, reinterpreting the ATLAS analysis which sets limits on
 theories of quark-WIMP effective interactions in terms of theories of
 photon-WIMP effective interactions (see the bottom of
 Fig.~\ref{diag}), working in an effective theory framework with a very
simple parameter space.

This class of interactions is of particular interest as collider
production of $\gamma+\missET$ via $\gamma^*/Z\rightarrow \gamma\chi\bar{\chi}$
is tied directly to the cross-section of putative monochromatic $\gamma$-ray
signals via $\chi\bar{\chi}\rightarrow \gamma\gamma$, allowing the
confrontation of LHC and Fermi-LAT data in the parameter space of our
new effective theory.  In this paper, we place  bounds from
$\gamma+\missET$ in the space of parameters that would
generate a signal at Fermi-LAT.

\subsection{Model}

We consider effective operators through which pairs of neutral stable
particles may couple to photons and possibly also the $Z$ boson.  We  consider operators where the DM
particles involved are scalars, as well as those in which they are fermions.

The most relevant (lowest-dimensional) operators involving scalar DM particles, $\phi$, are the dimension-6 operators:
 \bea
\label{eq:SMSinglet}
\mathcal{L}= \frac{1}{\Lambda_{B1,2}^2} ~\bar{\phi} \phi ~ \sum_i k_i
F_i^{\mu \nu} F^i_{\mu \nu} \nonumber \\+\frac{1}{\Lambda_{B3,4}^2} ~\bar{\phi} \phi ~ \sum_i k_i  F_i^{\mu \nu} \tilde{F^i_{\mu \nu}}
\eea
\noindent
where $F_i$, $i=1,2$ are the field strengths of the SM $U(1)$,
$SU(2)$ gauge groups.  Here and below, we label the effective
cut-offs of distinct operators $\Lambda_{Bi}$ or $\Lambda_{Ci}$ using the
notation of Ref.~\cite{Rajaraman:2012fu}. Note that in the second
operator, the dual field strength tensor appears.

Similarly, the most relevant operators involving fermionic DM, $\chi$,
are the dimension-7 operators:
\bea
\mathcal{L}= \frac{1}{\Lambda_{C1,2}^3} ~\bar{\chi} \chi ~ \sum_i k_i  F_i^{\mu \nu} F^i_{\mu \nu}+\nonumber\\\frac{1}{\Lambda_{C3,4}^3} ~\bar{\chi} \chi ~ \sum_i k_i  F_i^{\mu \nu} \tilde{F^i_{\mu \nu}}
\eea
and
\bea
\mathcal{L}= \frac{1}{\Lambda_{C5,6}^3} ~\bar{\chi} \gamma^5 \chi ~ \sum_i k_i  F_i^{\mu \nu} F^i_{\mu \nu}+\nonumber\\\frac{1}{\Lambda_{C7,8}^3} ~\bar{\chi} \gamma^5 \chi ~ \sum_i k_i  F_i^{\mu \nu} \tilde{F^i_{\mu \nu}},
\label{C56}
\eea
where here there are more operators than in the scalar case as the fermionic bilinears can have
different Lorentz structures. While the differences in collider limits on these operators are small 
(they differ only in their phase-space structure), the differences in their cosmic annihilation rates 
are substantial, as C1-C4 are velocity suppressed, $\sigv\sim \upsilon^2$, while C5-C8 are not. 

Given the form of the operators in Eqns.\ \ref{eq:SMSinglet}-\ref{C56} the couplings of the DM to various
pairs of electroweak gauge bosons are simply related by gauge symmetry:
\bea
g_{WW}&=&\frac{2k_2}{s_w^2 \Lambda^{2-3}} \\
g_{ZZ} &=& \frac{1}{4 s_w^2 \Lambda^{2-3}} \left(\frac{k_1 s_w^2}{c_w^2}+\frac{k_2 c_w^2}{s_w^2} \right) \\
g_{\gamma\gamma}&=&\frac{1}{4 c_w^2}\frac{k_1+k_2}{\Lambda^{2-3}} \\
g_{Z\gamma} &=& \frac{1}{2 s_w c_w \Lambda^{2-3}} \left(\frac{k_2}{s_w^2}-\frac{k_1}{c_w^2} \right),
\label{eq:prefactors}
\eea
where $s_w$ and $c_w$ are the sine and cosine of the weak mixing angle, respectively.

The parameters $k_1$ and $k_2$ control the relative couplings to
electro-weak guage bosons, but the fact that the couplings of pairs of DM particles to pairs of electro-weak gauge bosons are not all independent
will be very important here. The ratios of the couplings to the four possible two boson final states are simply
determined by two parameters ($\eg$, the $WW$ coupling can be turned off if $k_2=0$ while the $ZZ$ coupling is 
non-zero as long as either $k_1\neq 0$ or $k_2\neq 0$). The total cross-sections can be described in terms of 
three parameters: the ratios $k_{1,2}/\Lambda$ and the mass of the DM. As one moves around in this space the 
ratios of collider mono-boson production in various channels will change (along with the kinematics of such production,
$\eg$, the shape of the $\missET$ spectrum). There are also obviously regions of $k_{1,2}$ where interference between 
the underlying diagrams can completely suppress these interactions. This important fact will be reflected in our 
conclusions below.

In general one can write other operators involving the $WW$ or $ZZ$ gauge bosons at the same level in naive operator dimension\footnote{One 
can write even lower-dimensional operators, $\eg$, the Higgs Portal $|\phi|^2 V^2$ or $\bar\chi\chi V^2$ operators, but these aren't even $SU(2)\times U(1)$ 
invariant and so must clearly be related to the couplings of our operators in a UV model-dependent fashion.} 
but the relation between the coefficients of our operators Eqns.\ \ref{eq:SMSinglet}-\ref{C56} and of these other operators are not 
related by any symmetry of the Standard Model and so are UV model-dependent.

\subsection{Experimental Search}

The ATLAS experiment at the LHC has placed limits on  dark matter
production in the $\gamma+\missET$
channel~\cite{atlasphoton}, where the dark matter fields couple to quark initial states
and  the photon has been emitted as initial state radiation. These limits were derived from $4.6$ fb$^{-1}$ of data 
produced in $pp$ collisions at $\sqrt{s}=7$ TeV. The full 
selection is as follows:

\begin{itemize}
\item 1 photon, $p_{\textrm{T}}>150$ GeV
\item $\missET>150$ GeV 
\item $\le 1$ jet with $p_{\textrm{T}}>30$ GeV 
\end{itemize}

\begin{itemize}
\item $\Delta\phi(\gamma,\missET)>0.4$ 
\item $\Delta\phi(j_1,\missET)>0.4 $ 
\item No electrons (muons) with $p_{\textrm{T}}>$ 20 GeV and
  $|\eta|<2.47$ ($p_{\textrm{T}}>$ 10 GeV and
  $|\eta|<2.4$)
\end{itemize}

The results are consistent with the Standard Model expectation, as shown in Table~\ref{tab:atlasphoton}. 

\begin{table}
\begin{center}
\begin{tabular}{l|l l l}
\hline
\hline
Background source & \multicolumn{3}{c}{Events} \\
\hline
$Z(\rightarrow\nu\nu)\gamma$                      & $93$   & $\pm16$   & $\pm8$           \\
$Z/\gamma^{*}(\rightarrow\ell\ell)\gamma$ & $0.4$ & $\pm0.2$  & $\pm0.1$         \\
$W(\rightarrow\ell\nu)\gamma$                     & $24$   & $\pm5$    & $\pm2$           \\
$W/Z+$jets                                        & $18$   & \multicolumn{2}{c}{$\pm6$}   \\
Top                                               & $0.07$ & $\pm0.07$ & $\pm0.01$        \\
Diboson                                           & $0.3$  & $\pm0.1$  & $\pm0.1$         \\
$\gamma+$jets and multi-jet                       & $1.0$  & \multicolumn{2}{c}{$\pm0.5$} \\
\hline
Total & $137$ & $\pm18$ & $\pm9$ \\
\hline
Data  & \multicolumn{3}{c}{$116$} \\
\hline
\hline
\end{tabular}
\caption{\label{tab:atlasphoton} Breakdown of the number of data and background events as measured in the ATLAS mono-photon 
result~\cite{atlasphoton}. The first uncertainty is statistical and the second is systematic, except in the 
case of $W/Z+$jets, $\gamma+$jets, and multi-jet where the total uncertainty is quoted.}
\end{center}
\end{table}

Using the CLs method~\cite{cls1,cls2}, the ATLAS measurement constrains the number of non-Standard Model 
events to be $N<36$ at the 95\% confidence level. In order to reinterpret these results
in terms of interactions with electroweak bosons we must extract cross-section limits.
This can be done with the relation:
\begin{equation}
\sigma = \frac{N}{\mathcal{L}\times\epsilon}
\end{equation}
 where $\sigma$ is the cross section, $N$ is the number of events, 
$\mathcal{L}$ is the luminosity, $\epsilon$ is the total selection efficiency. 

\subsection{Signal Efficiency and Limits}
\label{colliders}
We generate events in this model using {\sc madgraph}5~\cite{madgraph}.
The efficiency for our signal events to survive the ATLAS event
selection is estimated by breaking the complete efficiency into two
parts: fiducial efficiency  of the
selection criteria ($\epsilon_{\textrm{fid}}$) and object
reconstruction efficiency $\epsilon_{\textrm{reco}}$.  The  fiducial efficiency  can be reliably estimated
using parton-level simulated event samples. The object
reconstruction efficiency depends on the details of the detector
performance, but is largely independent of the model.     We generate mono-photon ISR
events using the same configuration as the ATLAS analysis, measure the
fiducial efficiency for each operator, and use the reported ATLAS total efficiences to deduce the
object reconstruction efficiency.  This allows us to estimate the
total efficiency for our new signal events.

\begin{figure}
\includegraphics[width=0.45\linewidth]{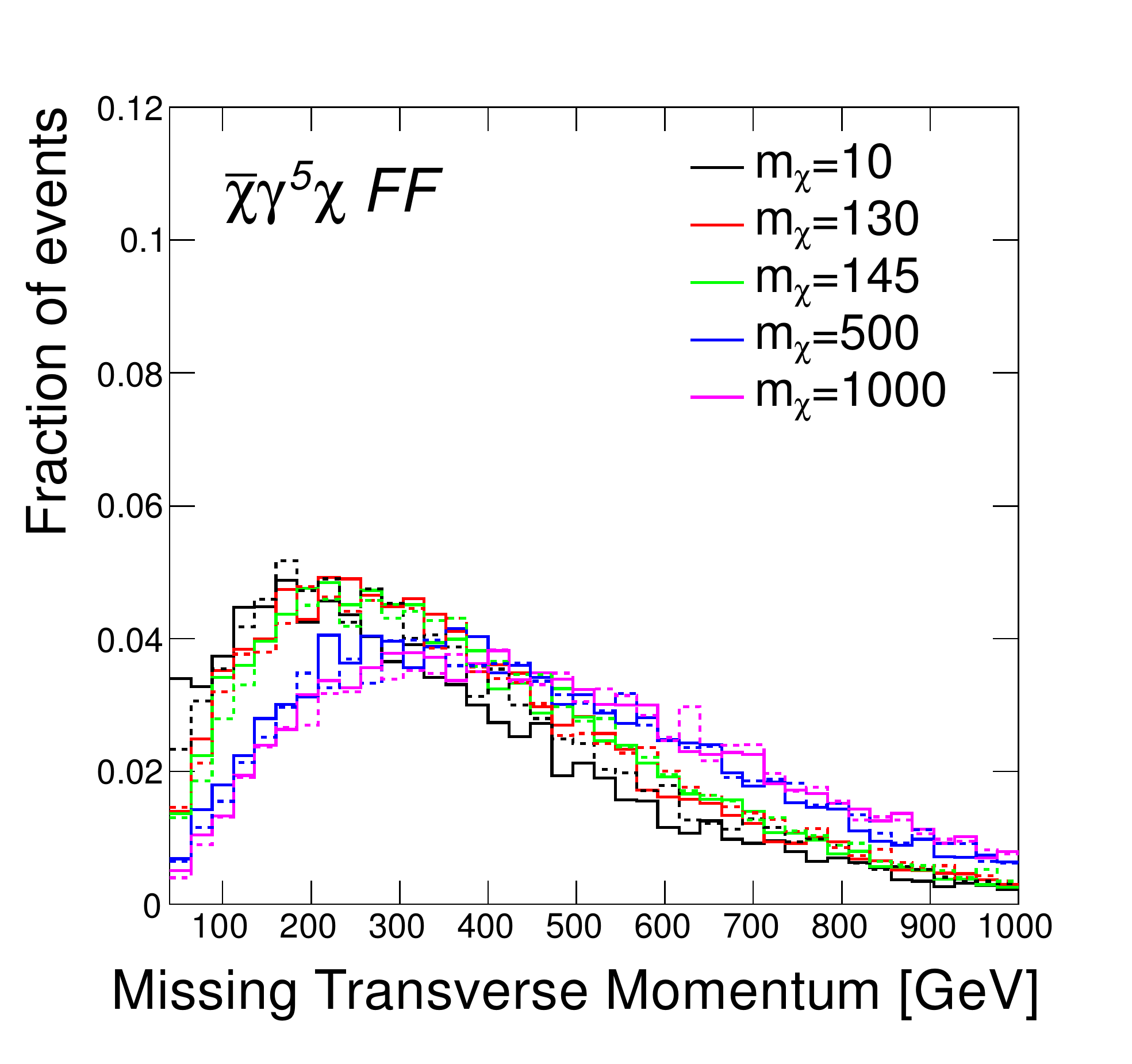}
\includegraphics[width=0.45\linewidth]{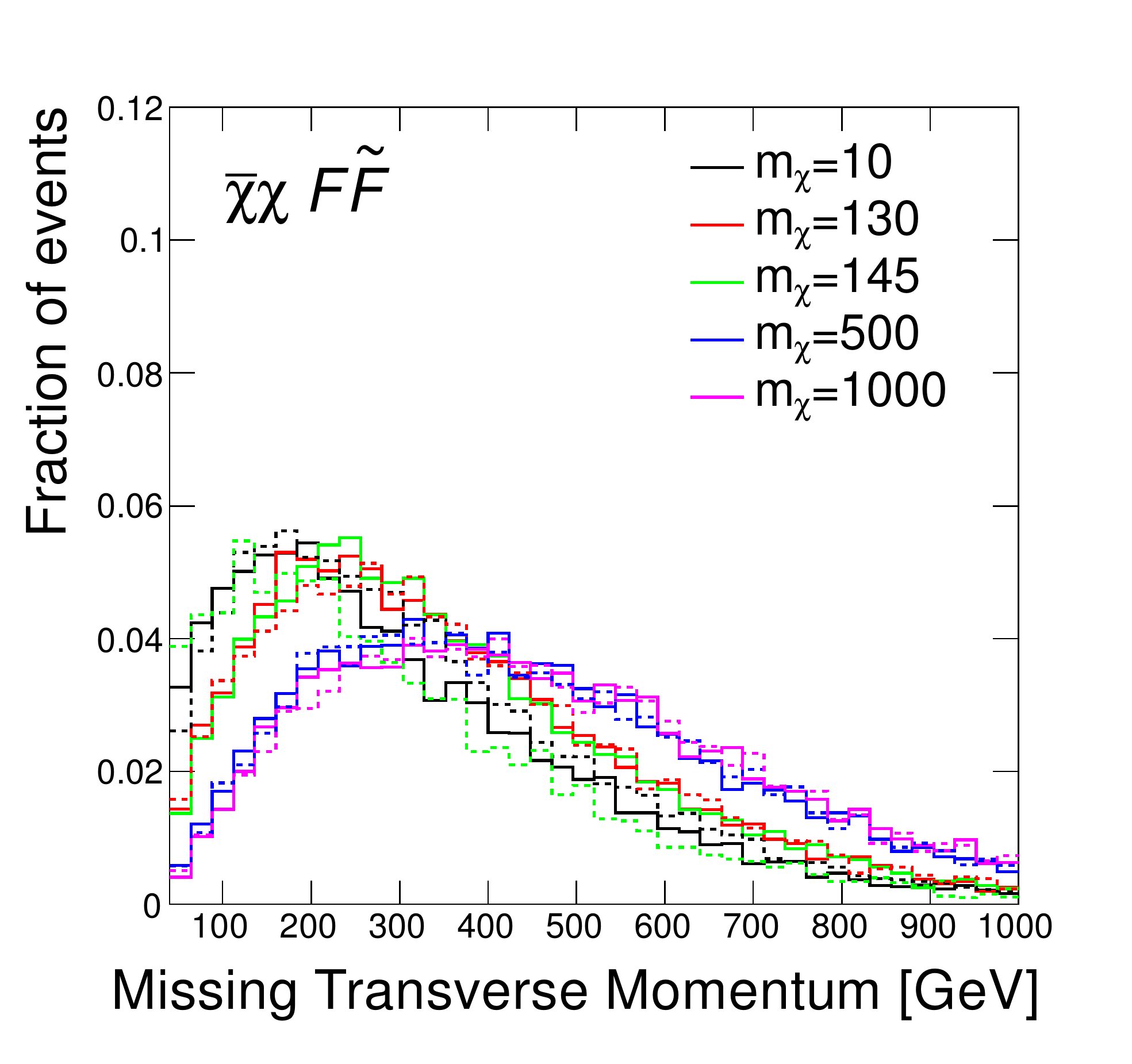}
\includegraphics[width=0.45\linewidth]{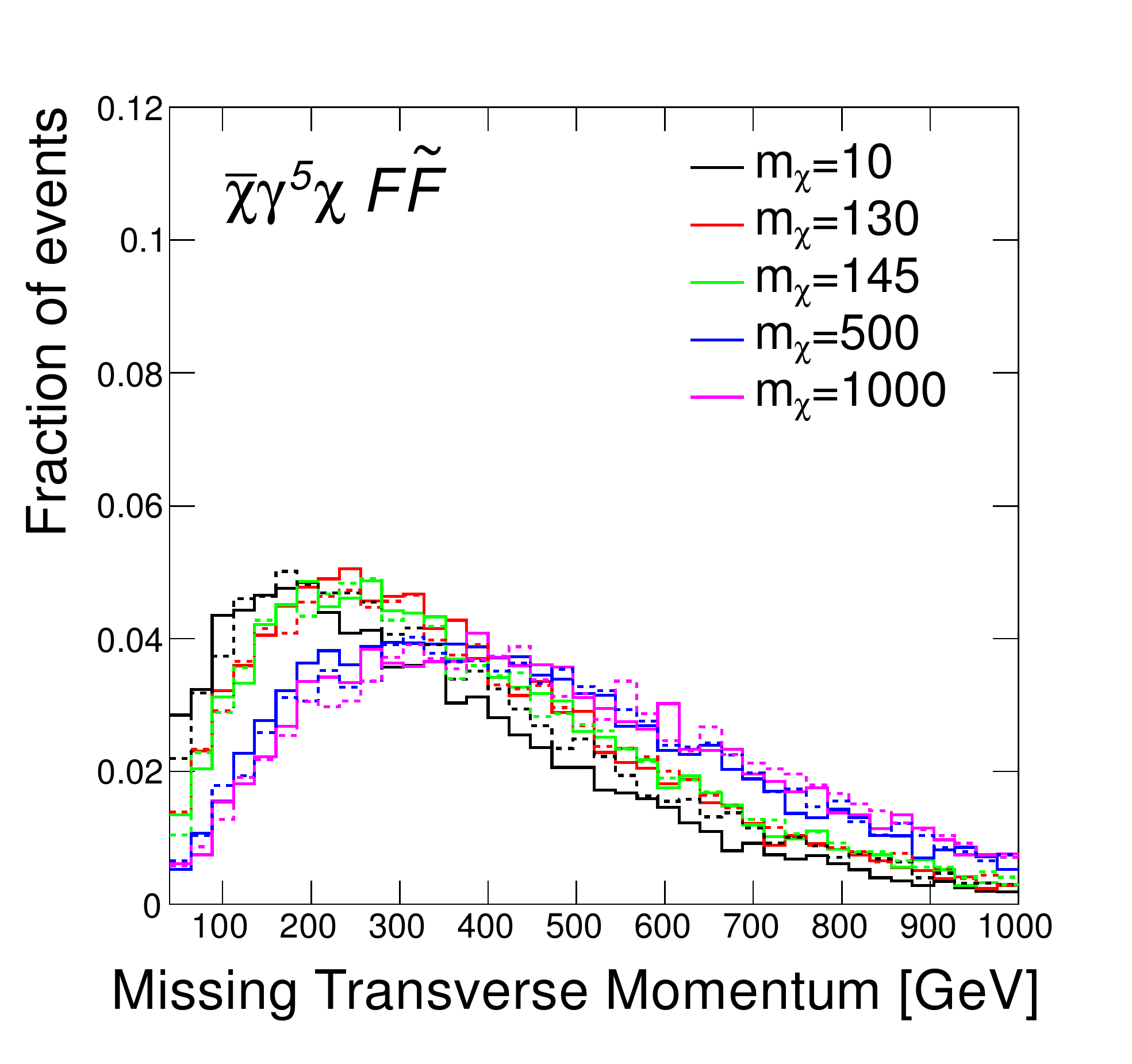}
\includegraphics[width=0.45\linewidth]{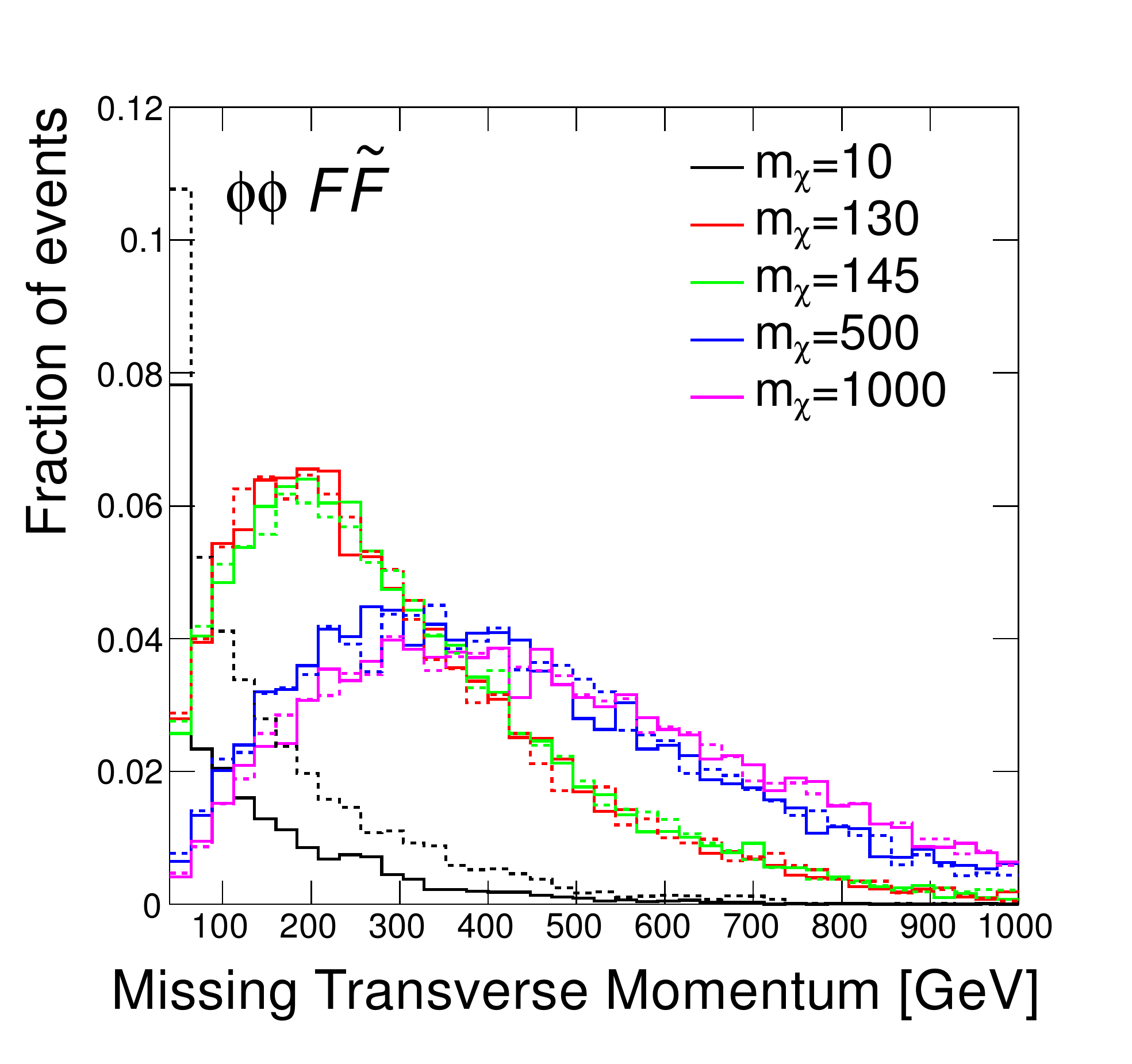}
\includegraphics[width=0.45\linewidth]{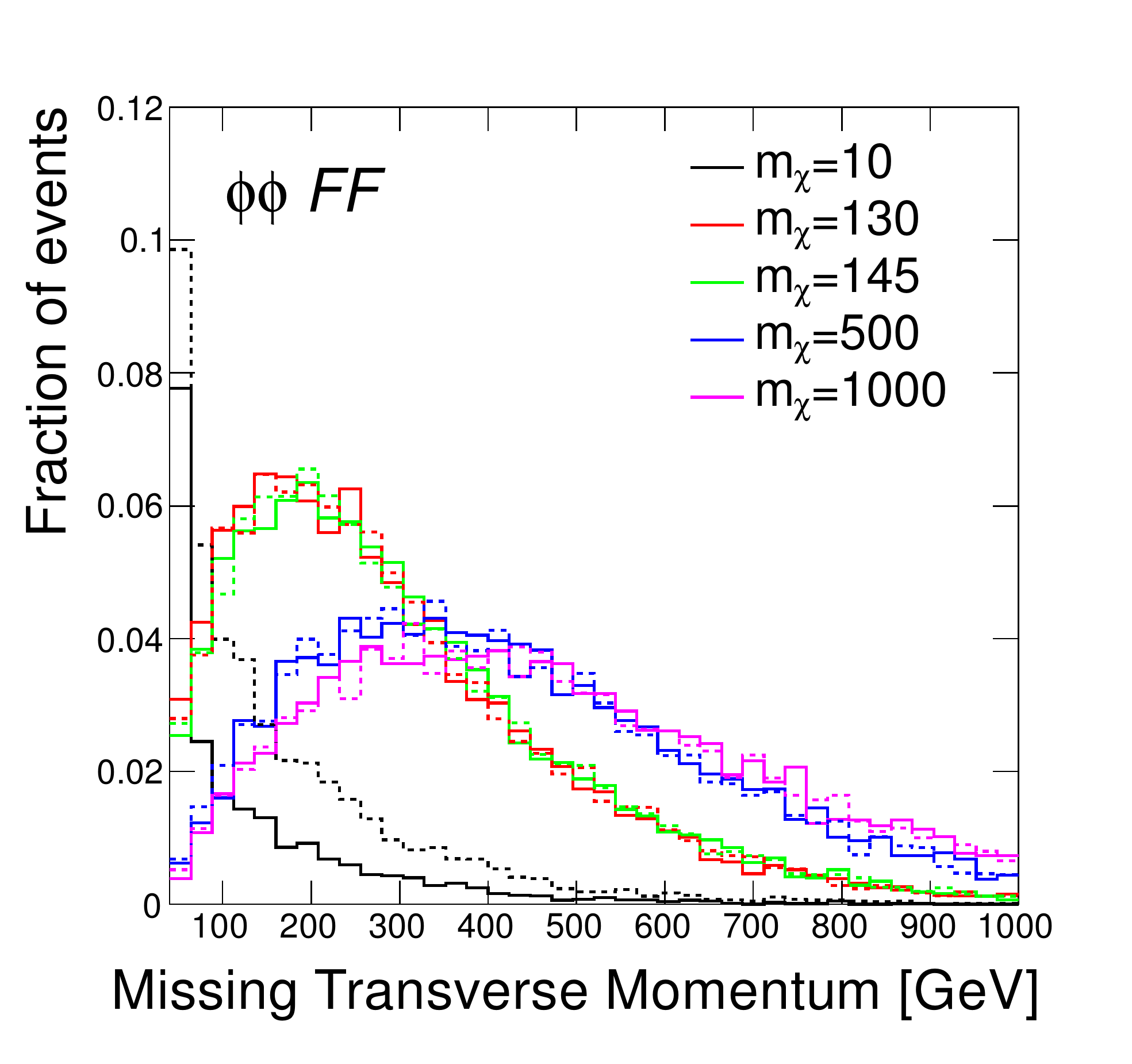}
\includegraphics[width=0.45\linewidth]{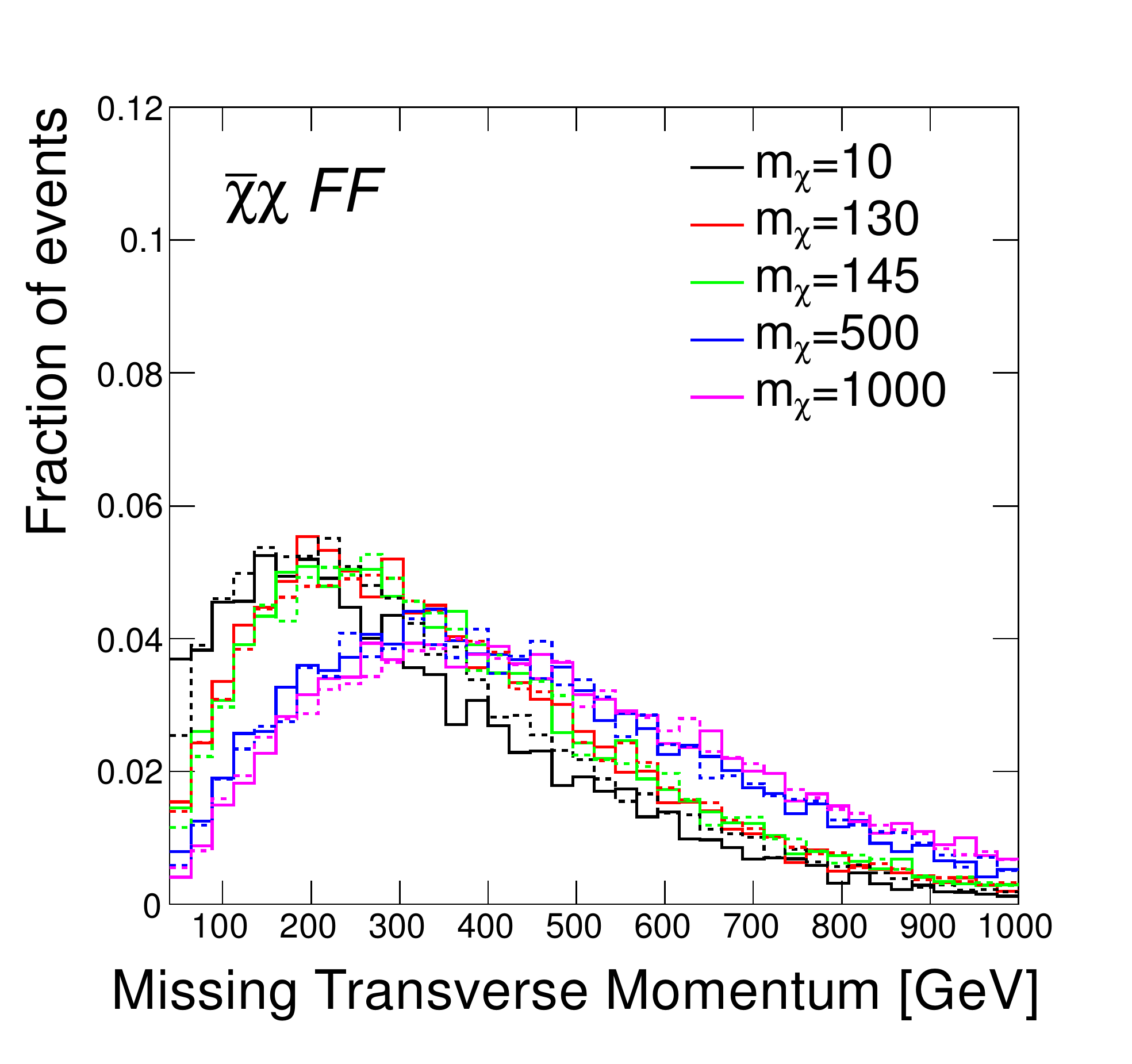}
\caption{ Distributions of $\missET$ in simulated $\gamma+\missET$
  events in $pp$ collisions at the
  LHC for several choices of  $m_\chi$
  and $k_1,k_2 = 0.5,0.5$ (solid) or $1,0$ (dashed).}
\label{fig:met}
\end{figure}

The critical kinematic quantity is the missing transverse
momentum. Figure~\ref{fig:met} shows the distributions for a few
choices of $k_1,k_2$ and $m_\chi$.

\begin{figure}
\includegraphics[width=0.45\linewidth]{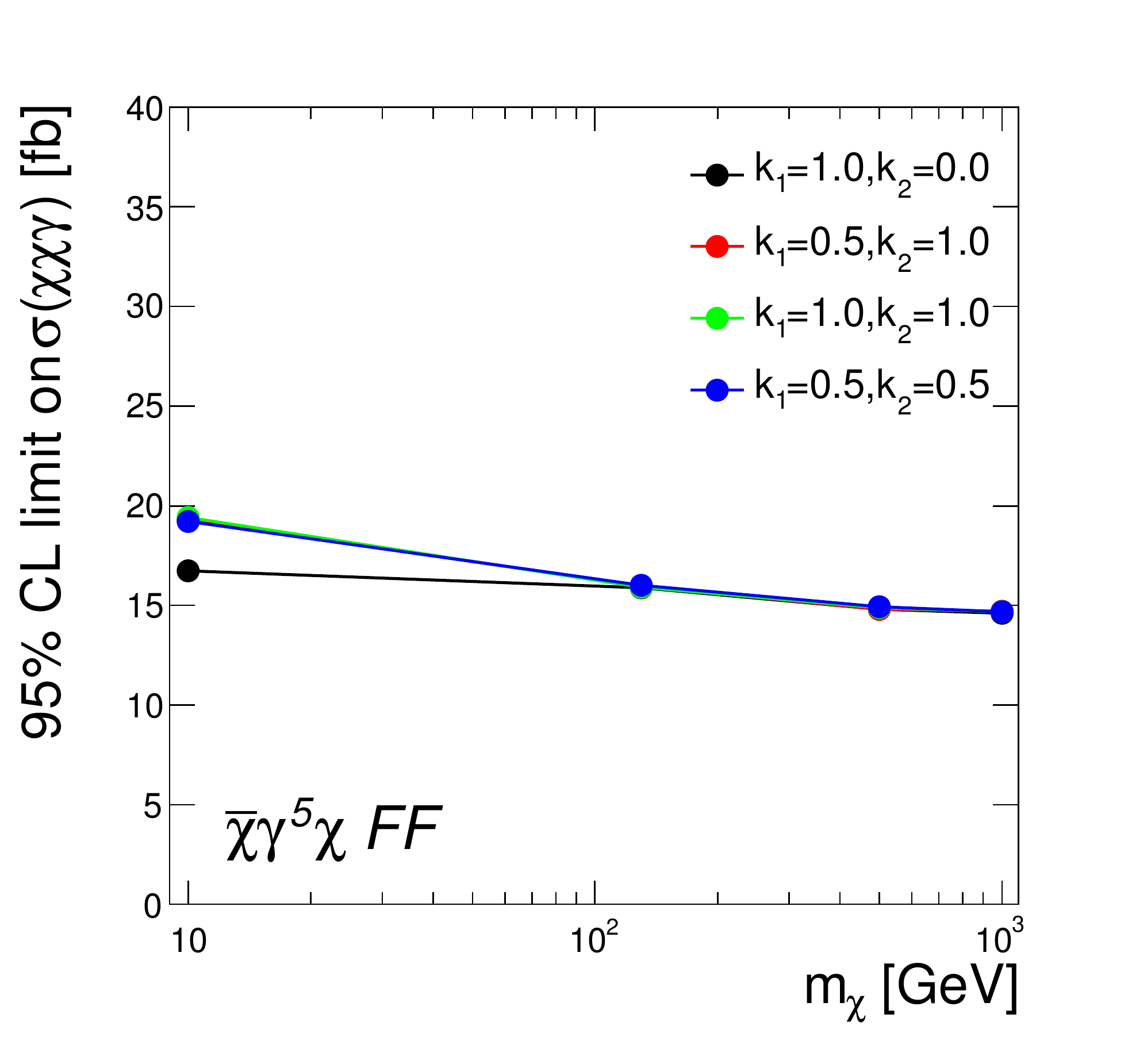}
\includegraphics[width=0.45\linewidth]{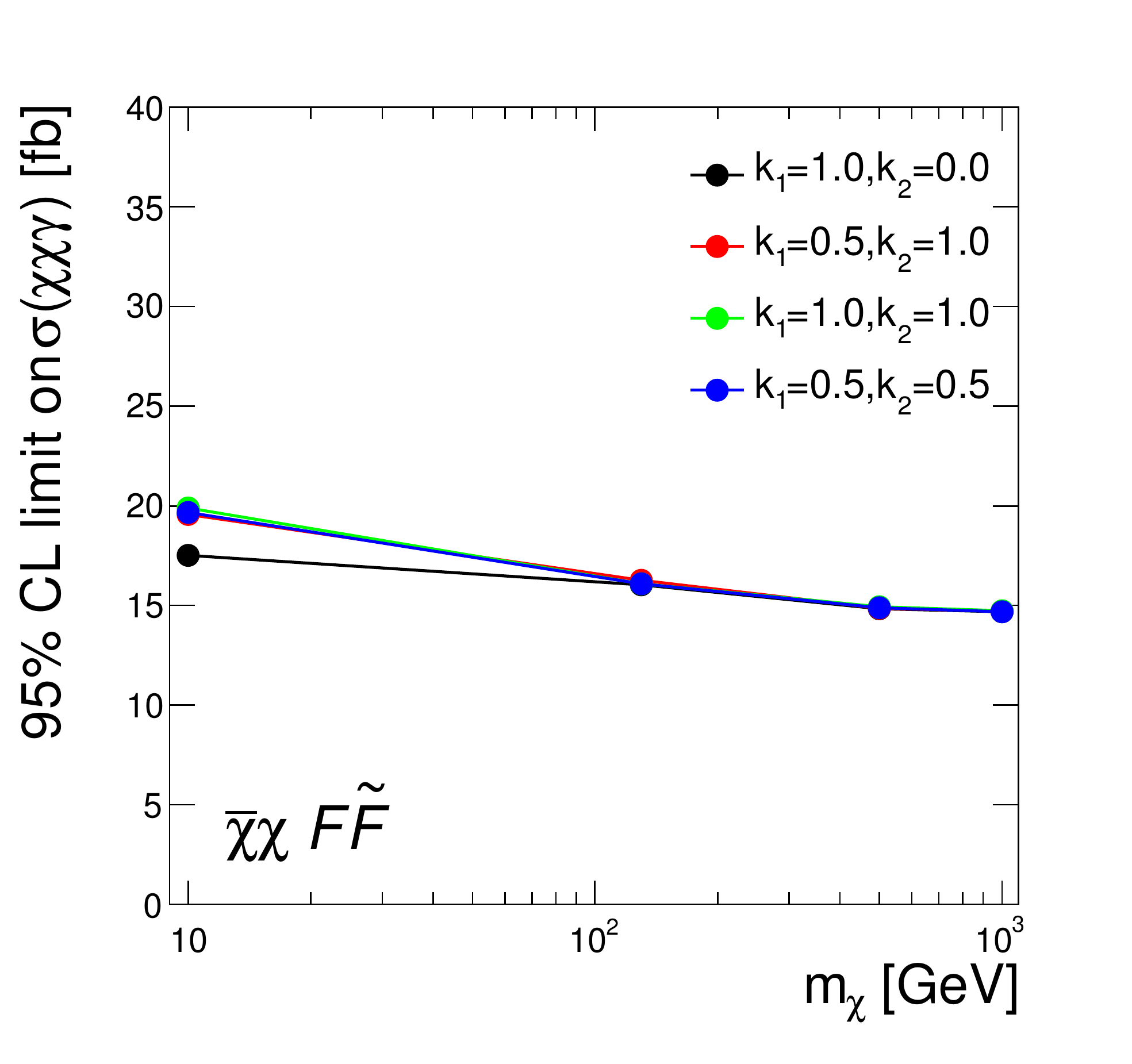}
\includegraphics[width=0.45\linewidth]{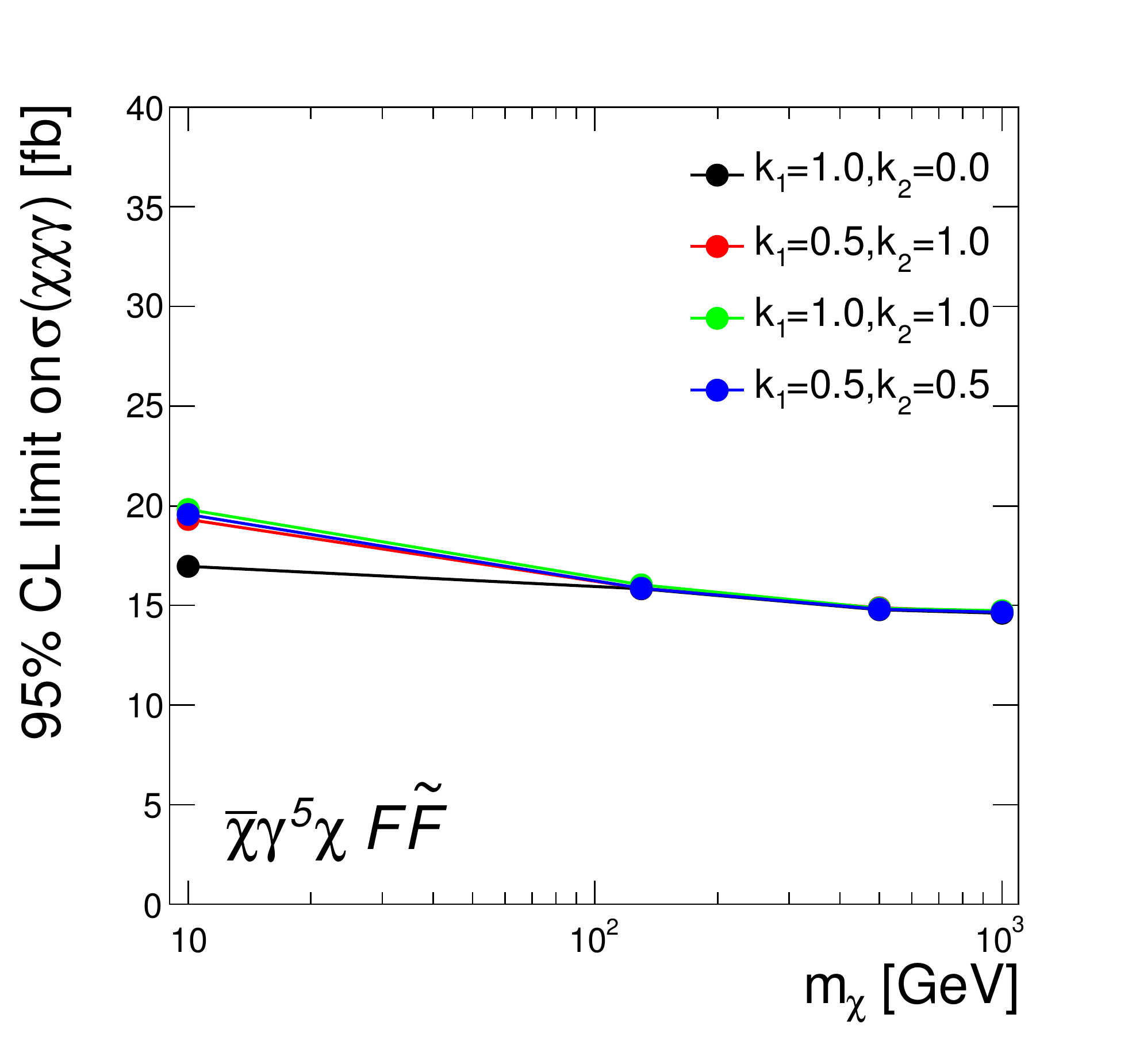}
\includegraphics[width=0.45\linewidth]{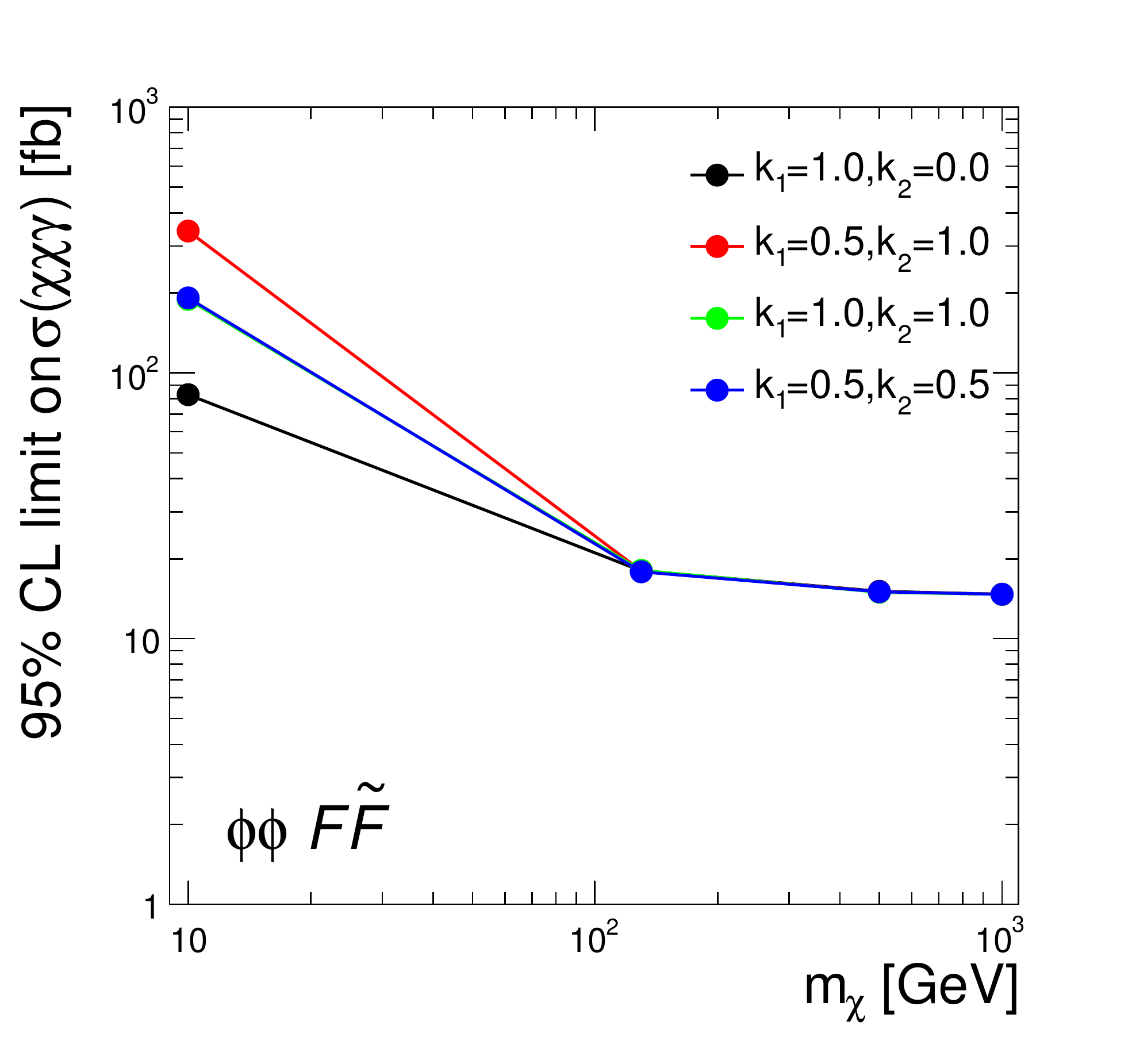}
\includegraphics[width=0.45\linewidth]{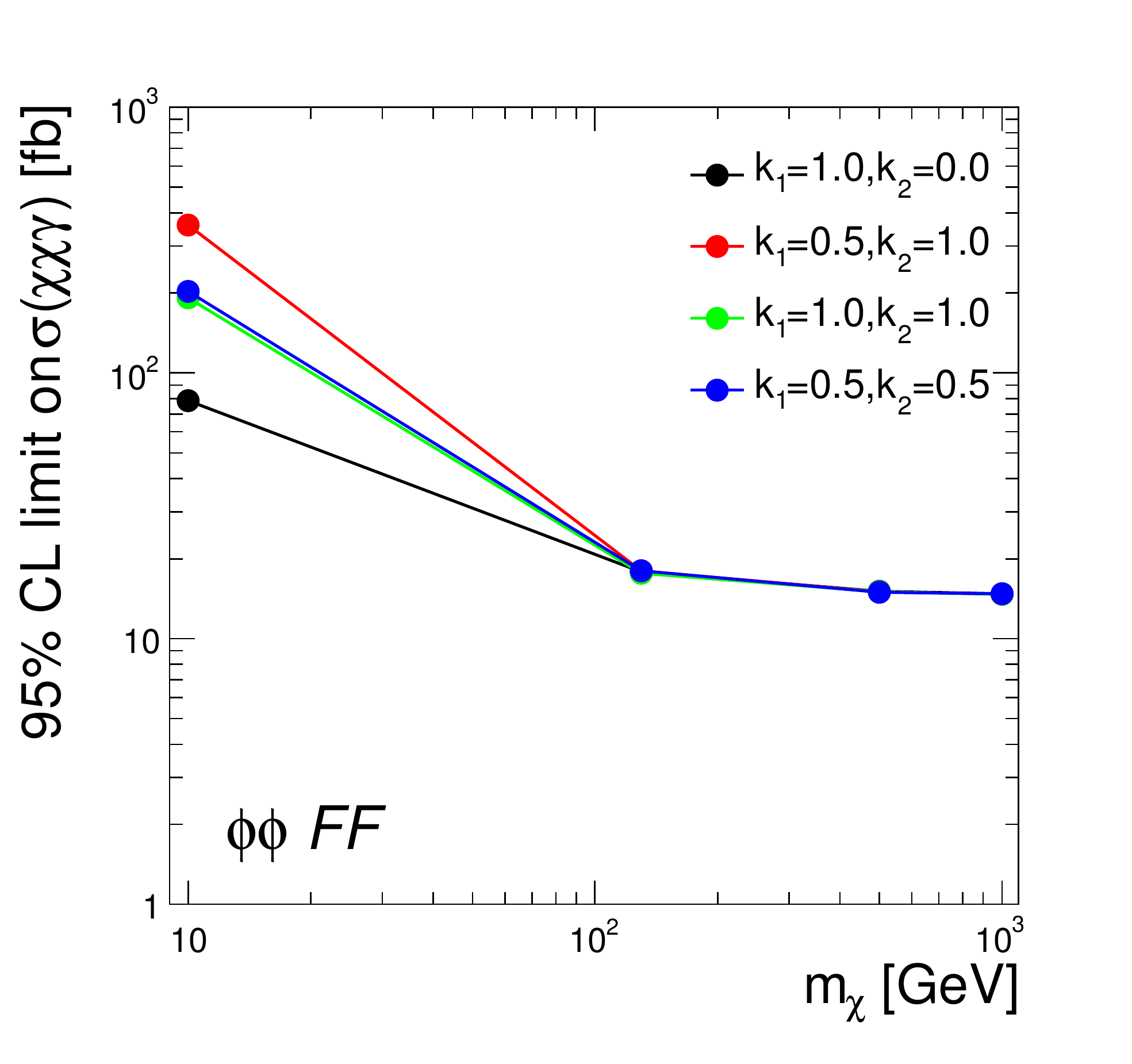}
\includegraphics[width=0.45\linewidth]{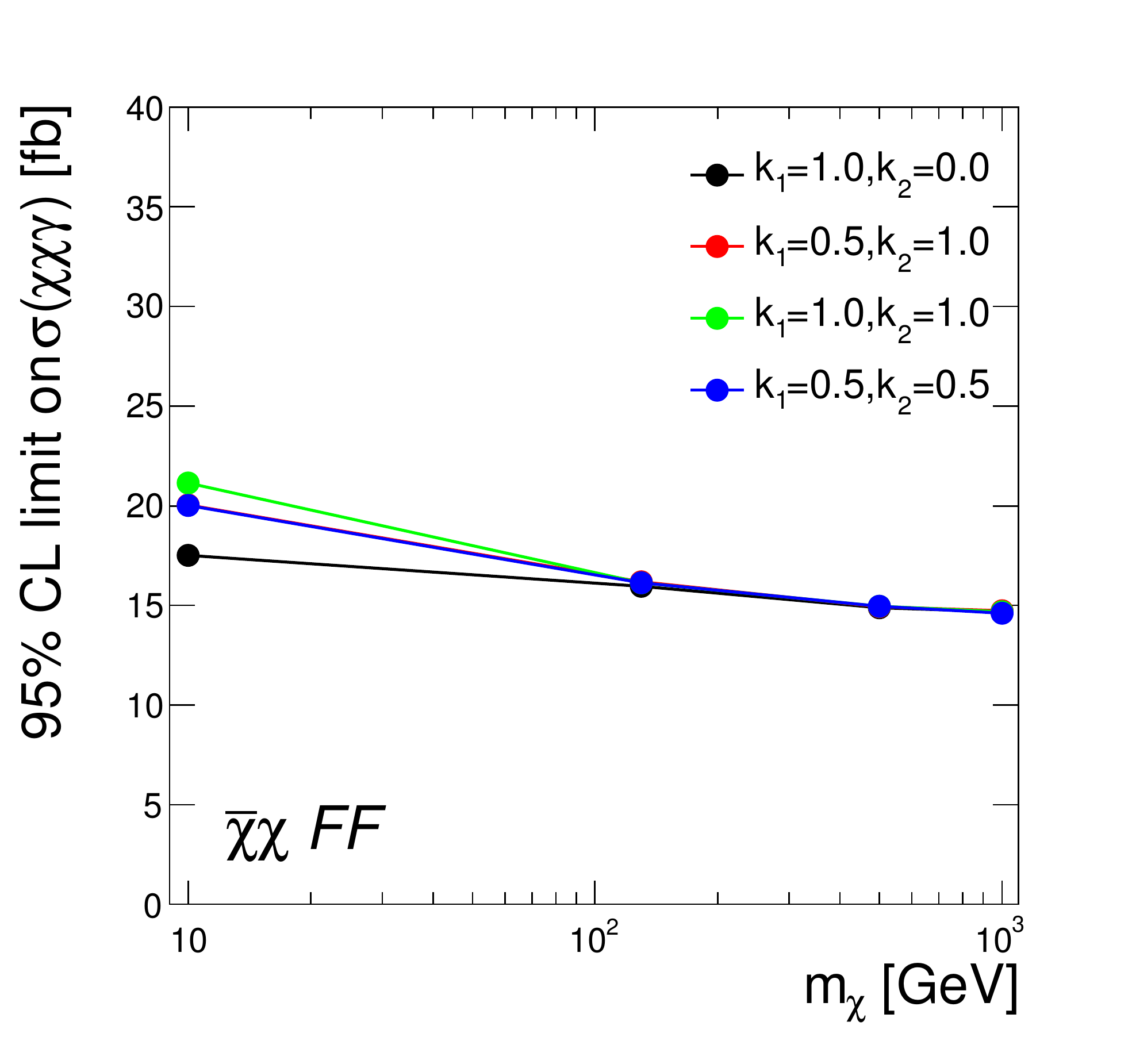}
\caption{ Limits on $\sigma(pp\rightarrow \gamma+\missET$ for several
  values of $m_\chi$ and $k_1,k_2$.}
\label{fig:lim_mchi}
\end{figure}

\begin{figure}
\includegraphics[width=0.45\linewidth]{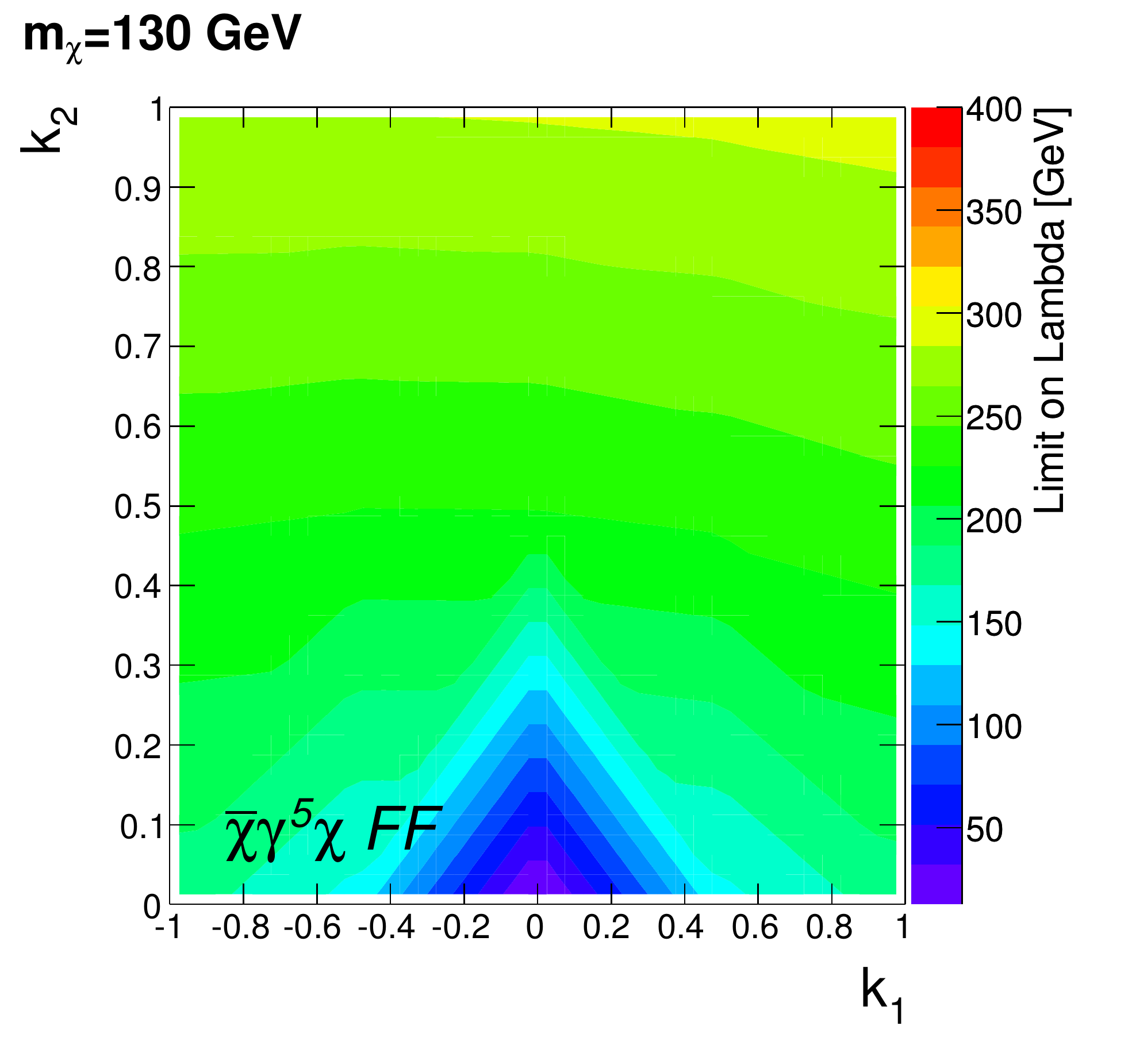}
\includegraphics[width=0.45\linewidth]{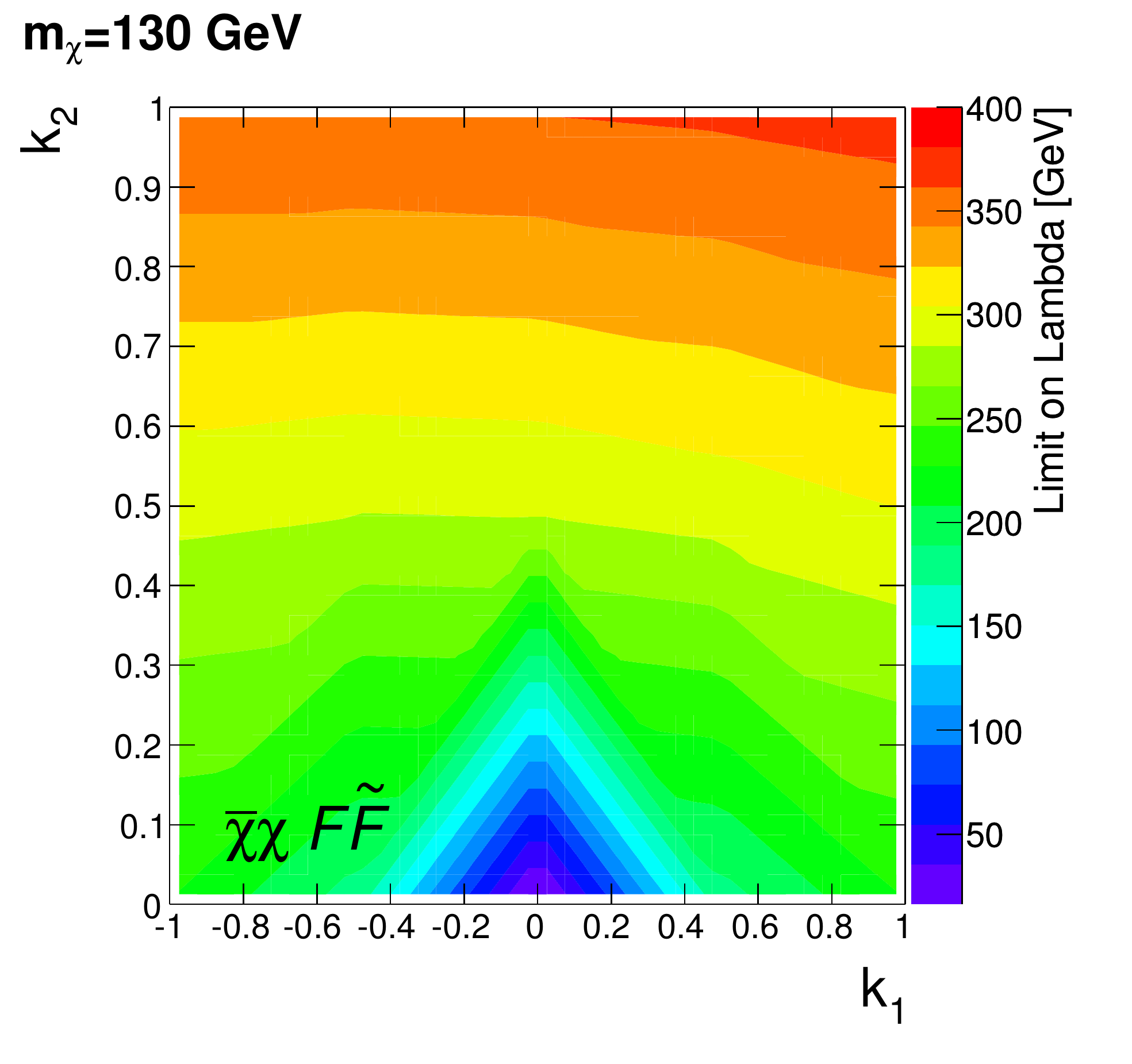}
\includegraphics[width=0.45\linewidth]{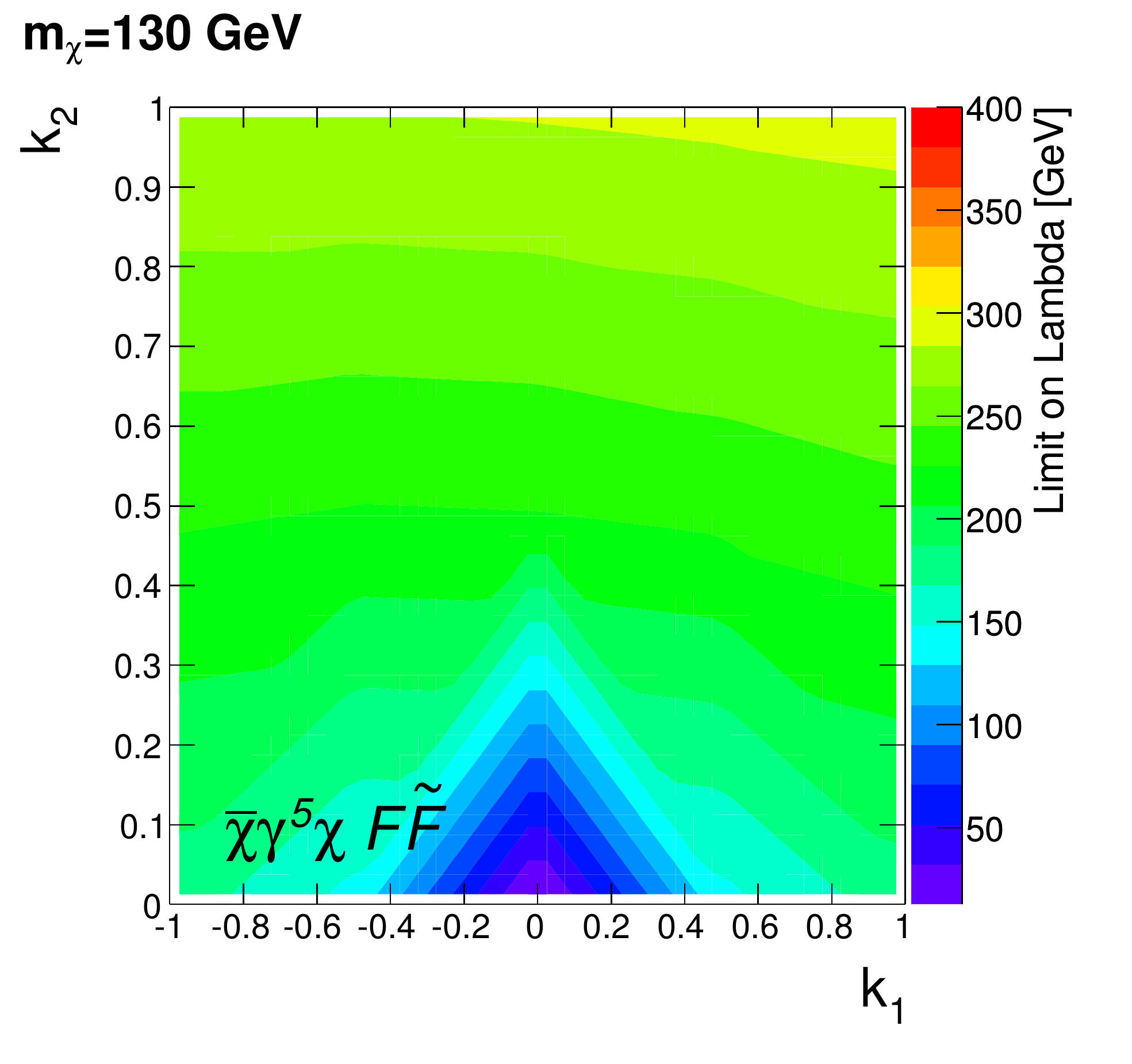}
\includegraphics[width=0.45\linewidth]{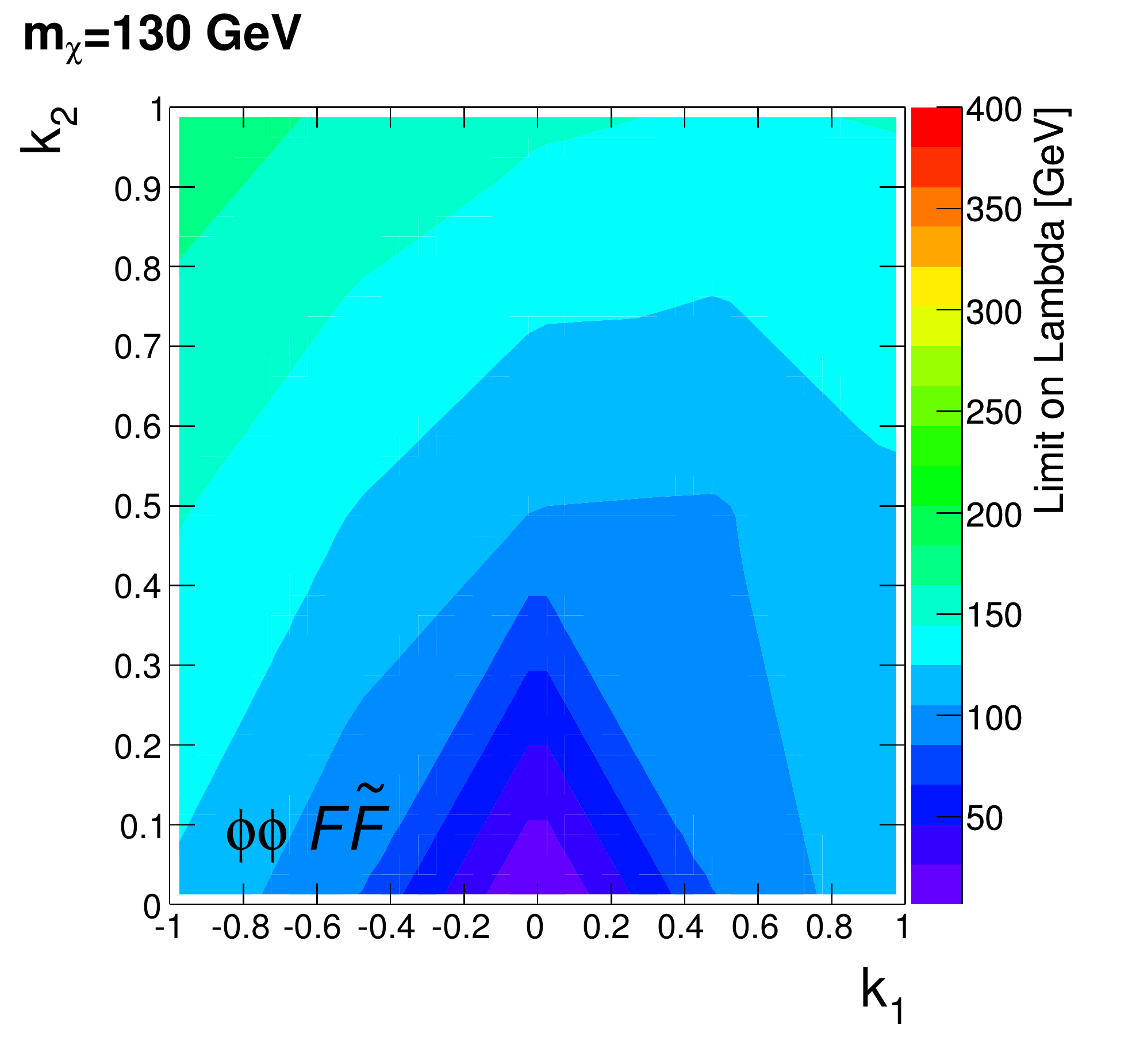}
\includegraphics[width=0.45\linewidth]{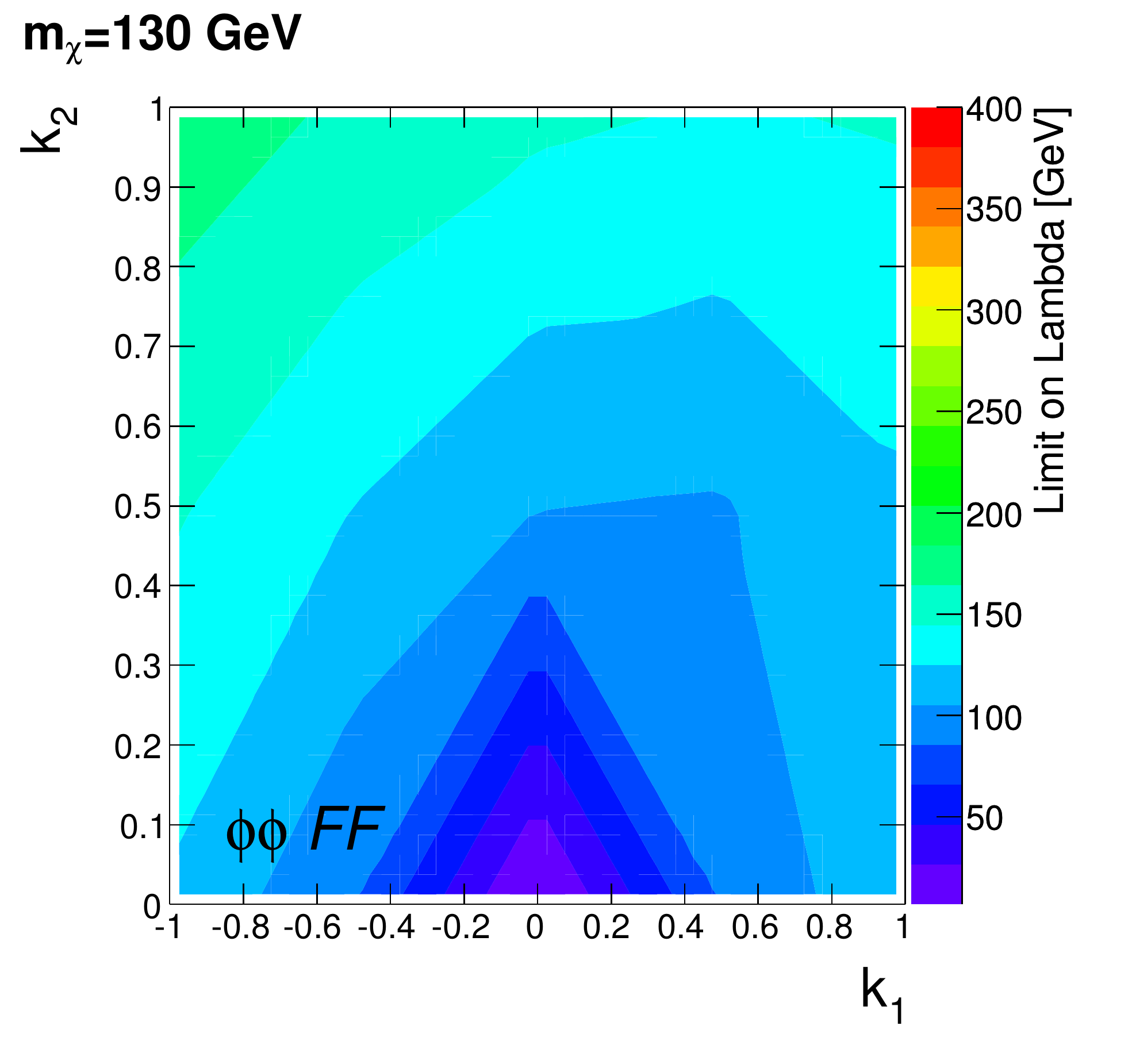}
\includegraphics[width=0.45\linewidth]{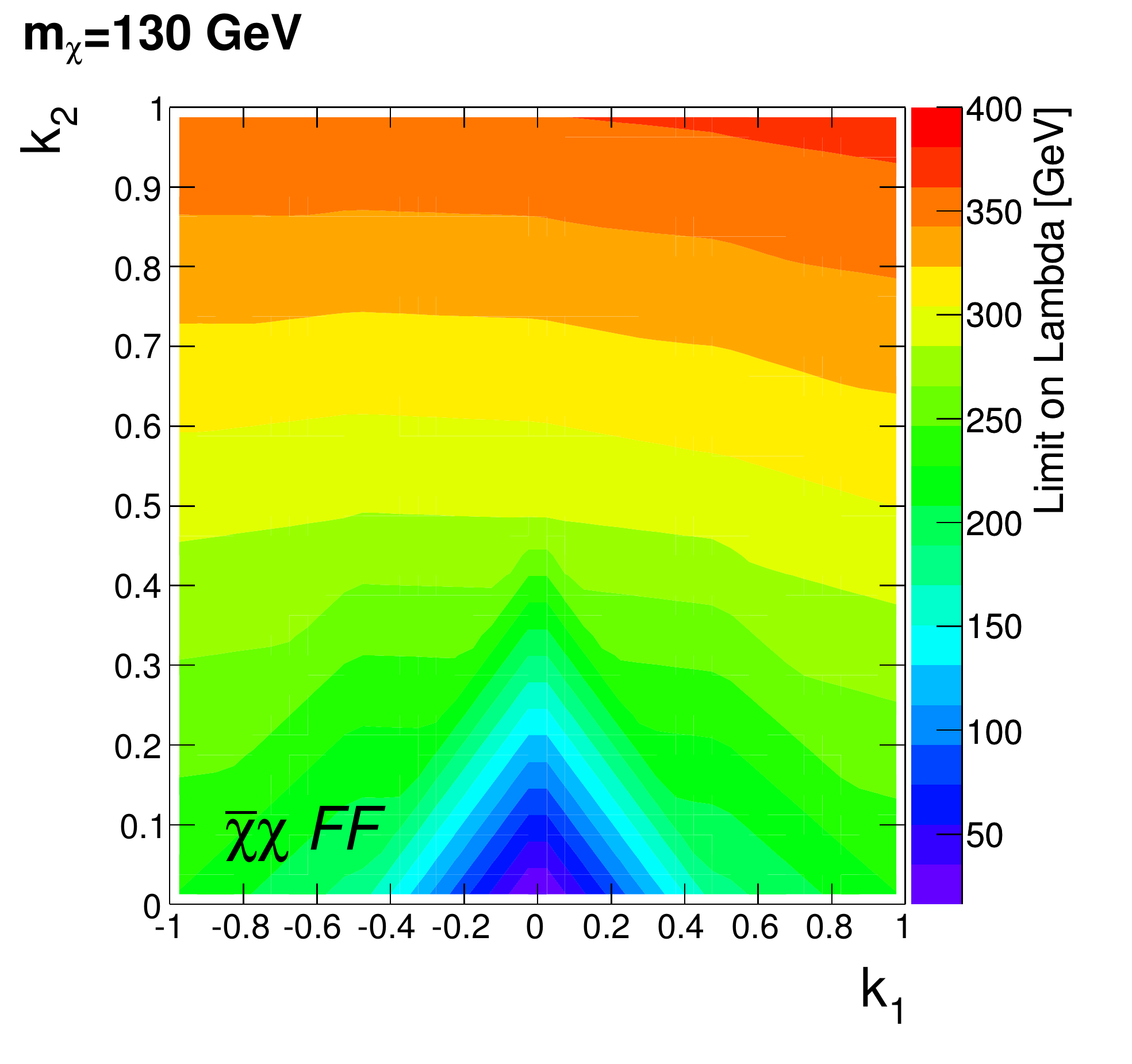}
\caption{ Limits on $\Lambda$ from $\gamma+\missET$ events at the LHC,
  with $m_\chi=130$ GeV as a function of $k_1,k_2$.}
\label{fig:lim_lambda}
\end{figure}

\begin{figure}
\includegraphics[width=0.45\linewidth]{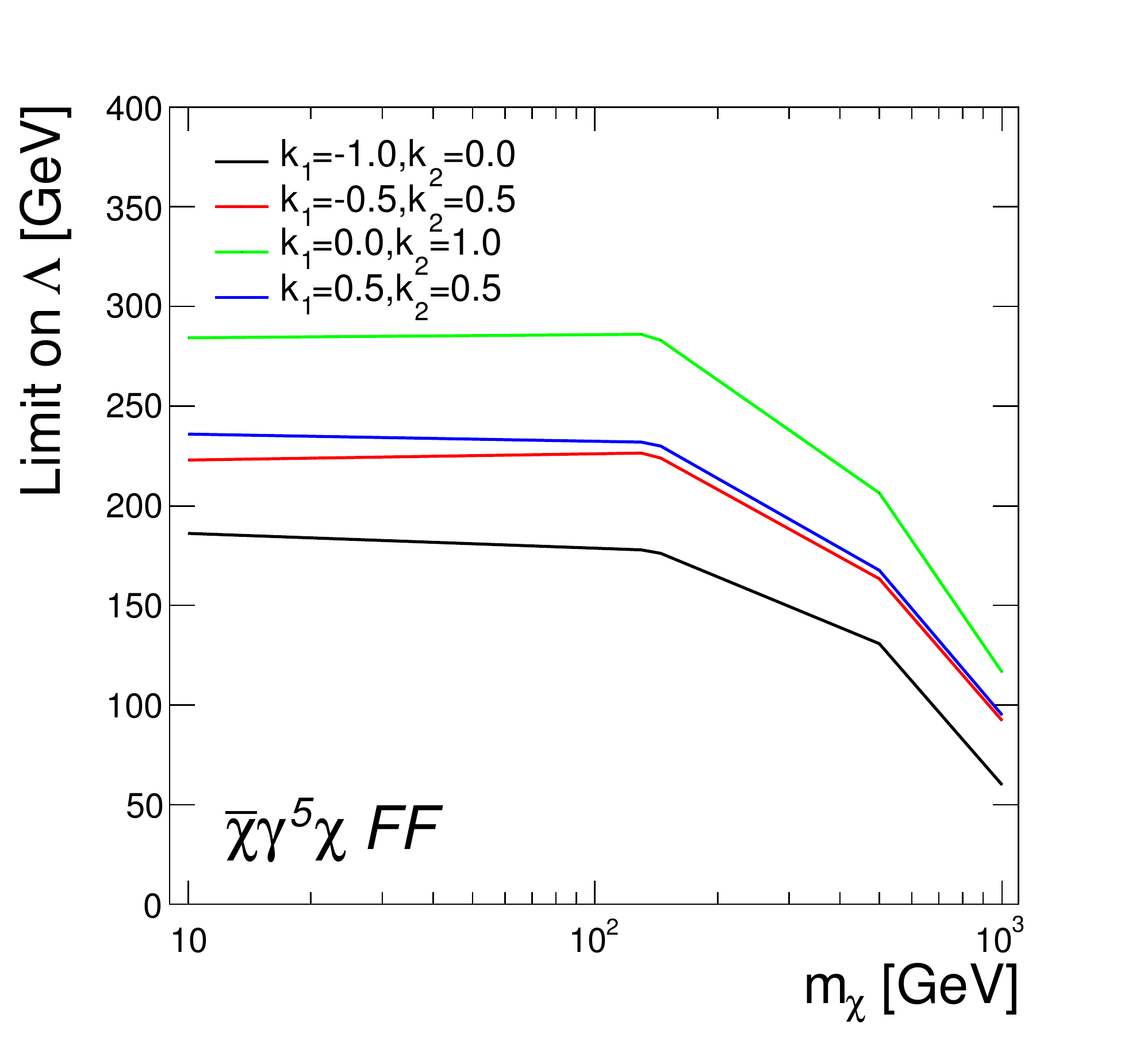}
\includegraphics[width=0.45\linewidth]{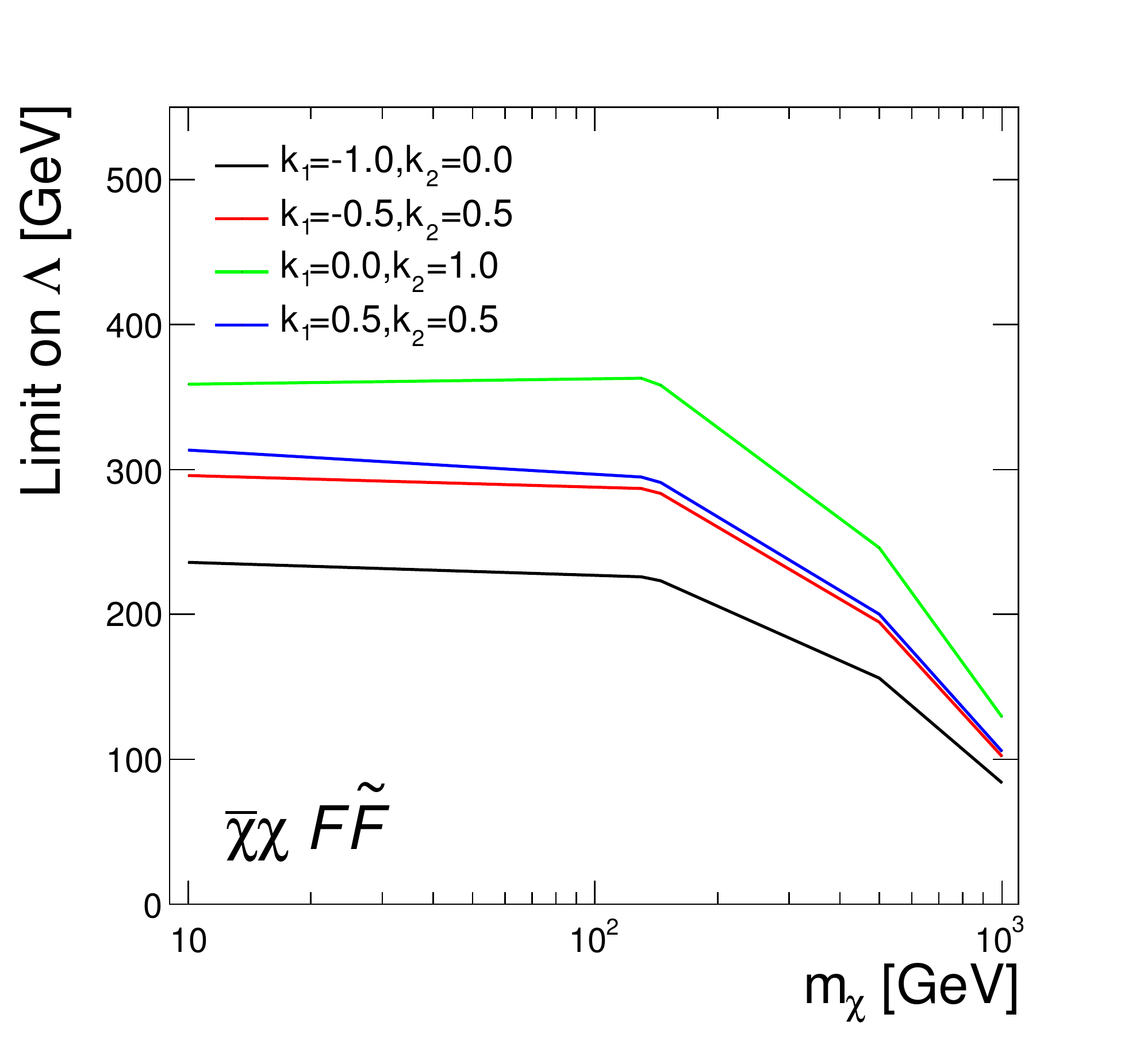}
\includegraphics[width=0.45\linewidth]{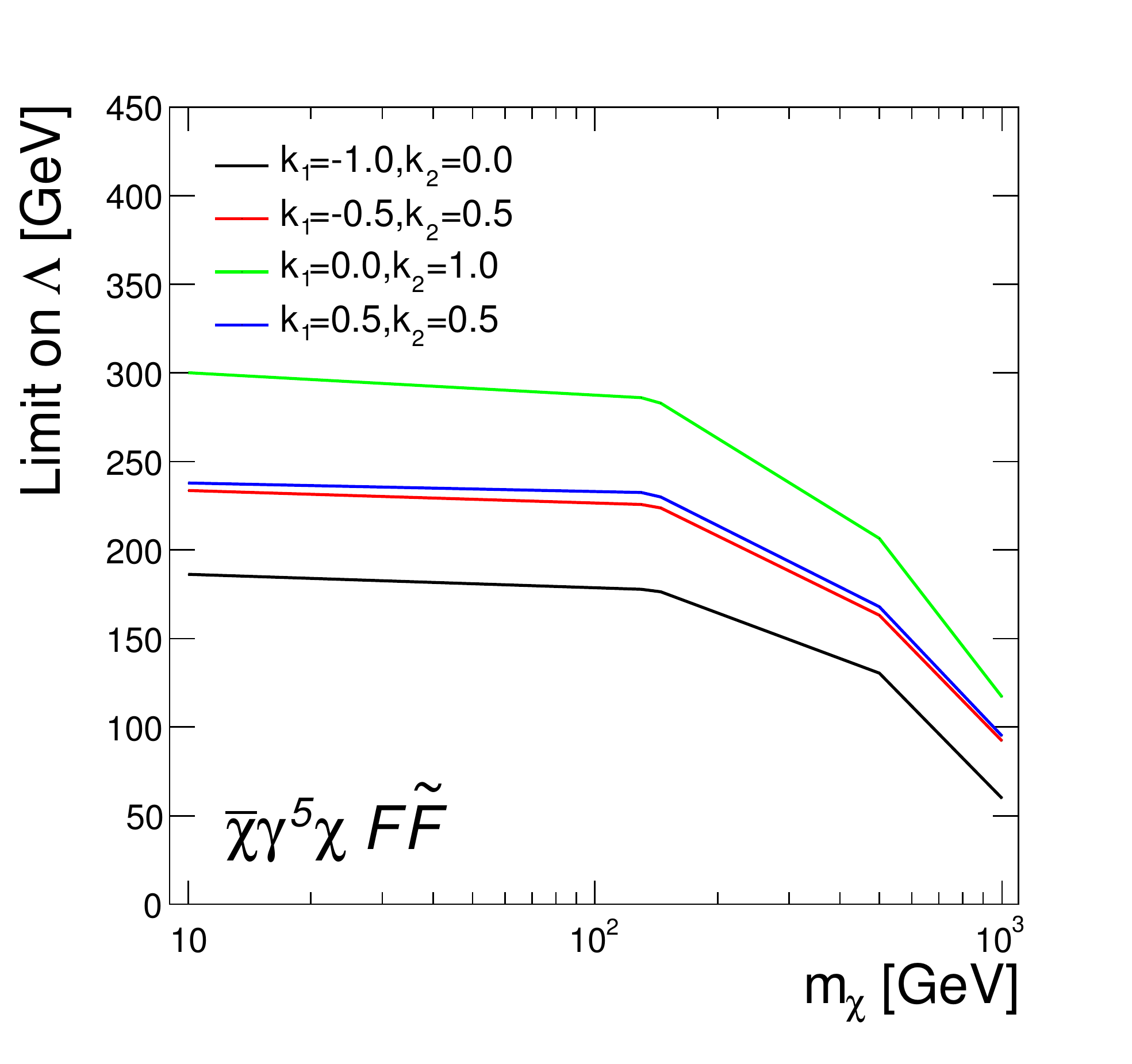}
\includegraphics[width=0.45\linewidth]{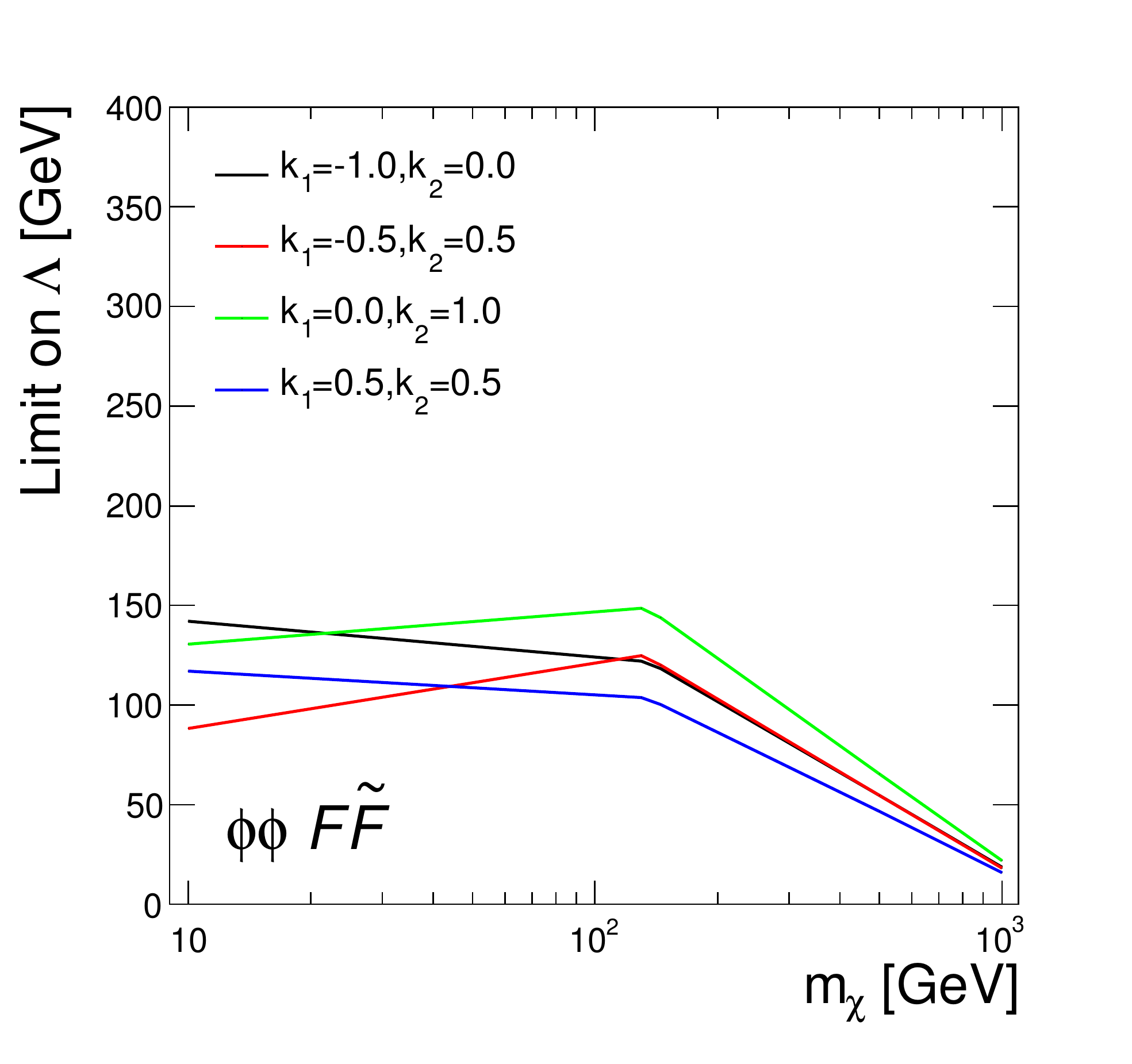}
\includegraphics[width=0.45\linewidth]{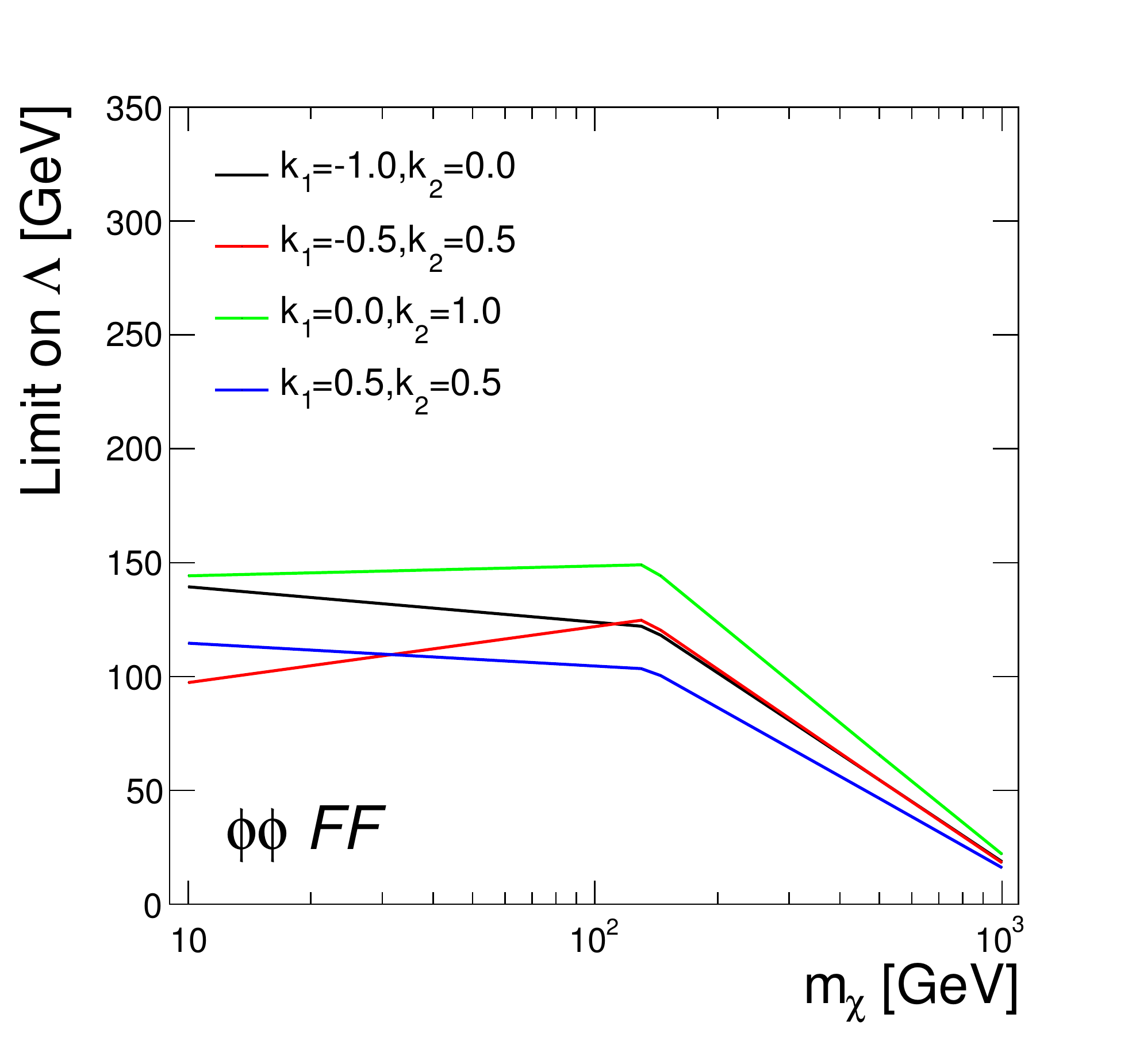}
\includegraphics[width=0.45\linewidth]{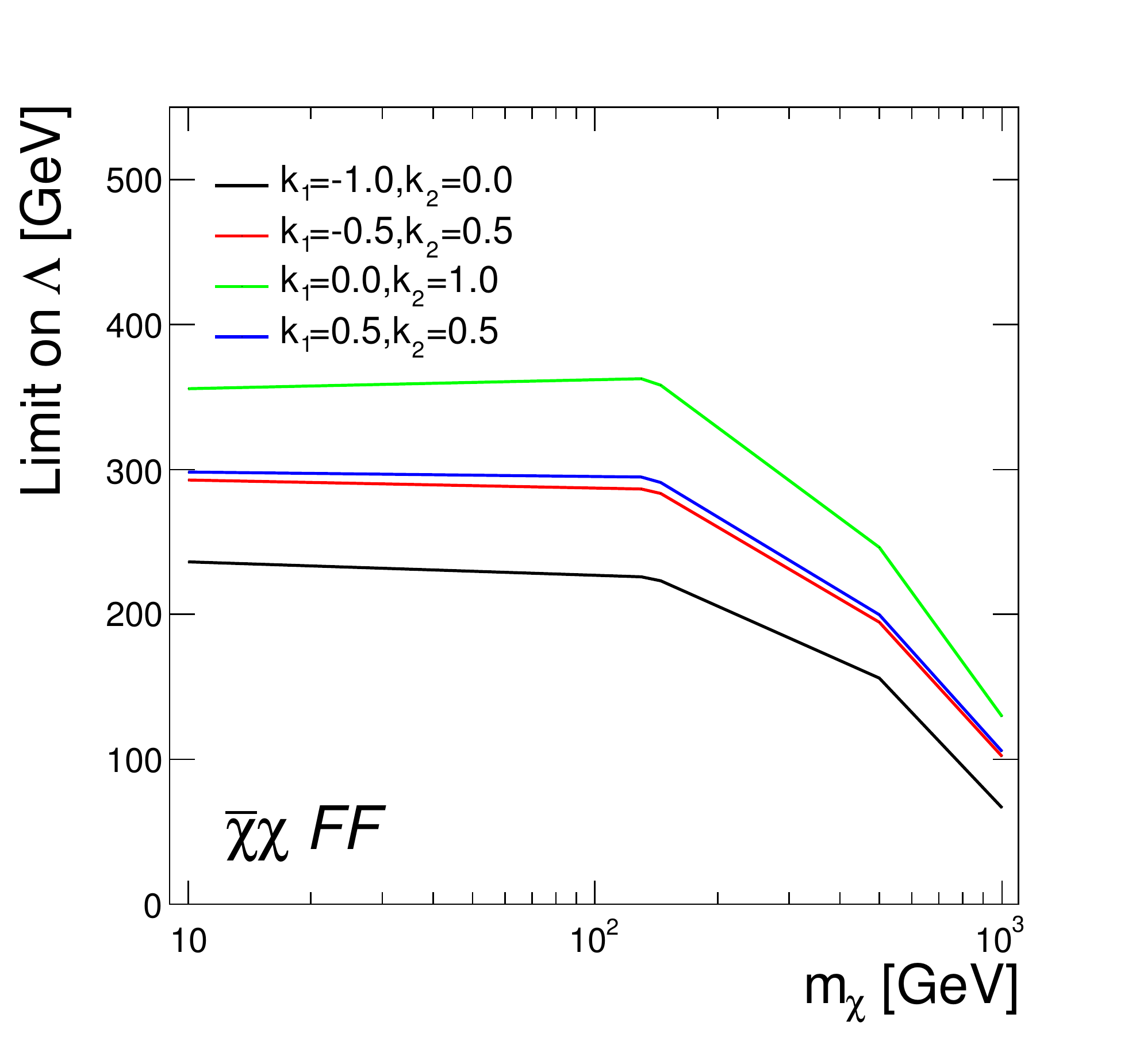}
\caption{ Limits on $\Lambda$ as a function of $m_\chi$ for
  several choices of $k_1,k_2$.}
\label{fig:lim_lambda2}
\end{figure}

Limits on the cross section are shown as a function of $m_\chi$ for
several choices of $k_1,k_2$ in Fig.~\ref{fig:lim_mchi}. As the
cross section depends on the suppression scale $\Lambda$,
limits on the cross section can be translated into limits on
$\Lambda$, see Figs~\ref{fig:lim_lambda}, ~\ref{fig:lim_lambda2}.

We observe (Fig.~\ref{fig:lim_mchi}) that the limits on light fermionic $\chi$ are much tighter than those
on light scalar $\phi$ DM, a feature that is obviously due to the differences in 
$\missET$ spectra (Fig.~\ref{fig:met}). It is not hard to understand these differences as, in the limit of 
massless $\chi$, the fact that the fermionic operators are dimension-7 and the scalar operators are 
dimension-6 means that the cross-sections in the fermionic case must scale with a higher power of 
the momenta involved\footnote{Terms that don't scale with momenta are much smaller, $\mathcal{O}(m^2_{\chi})/s$, $\ie$, 
they are ``helicity suppressed.''}, and hence the photon $p_T$. The resulting cross-section is relatively suppressed as $p_T\to 0$
and is enhanced compared to the scalar case in the large $p_T$ tail.

\subsection{Gamma-Ray Lines from our Operators}

\subsubsection{Line Rates}

For fermionic dark matter ($\chi$), the only operators that will give
sizeable annihilation rates to the $\gamma\gamma$ and $\gamma Z$ final
states are those which are not velocity suppressed ($\bar{\chi}\chi\sim\upsilon^2$, with $\upsilon\sim10^{-3}$). The relevant
operators here are C5-C8. In the case of scalar dark metter ($\phi$), none of the operators mentioned above, B1-B4, are suppressed.

\begin{figure}[hbtp]
    \centering
    \includegraphics[width=\linewidth]{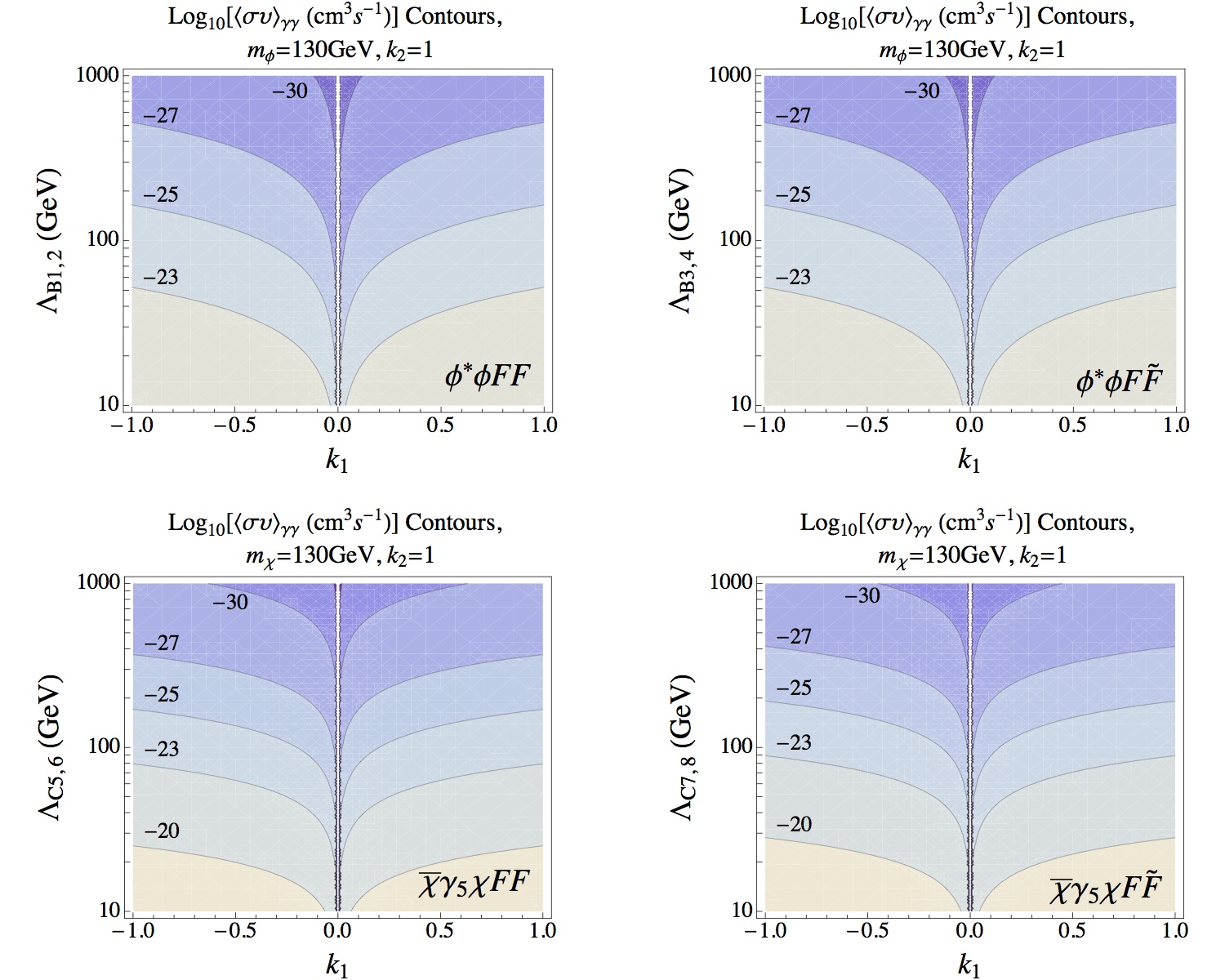}
    \caption{Annihilation rates for our (four unsuppressed) operators. We fix $m_{DM}=130\gev$ and $k_2=0$ here for illustration.}
    \label{anns}
  \end{figure}

Annihilation rates are straightforward to calculate for these operators, see $\eg$, Ref.~\cite{Rajaraman:2012fu} for a recent accounting of such calculations. In
terms of our parameterization we find:
\begin{eqnarray}
\langle\sigma\upsilon\rangle^{\gamma\gamma}_{B1,2}&=&\frac{ 2m_{\chi}^2 }{ \pi \Lambda_s^4 }\left( k_1 c_w^2 + k_2 s_w^2 \right)^2\\
\langle\sigma\upsilon\rangle^{\gamma Z}_{B1,2}&=&\frac{ 3(4m_{\chi}^2-m_{Z}^2)^3 c_w^2 s_w^2 }{ 64\pi m_{\chi}^4 \Lambda_s^4 }\left(k_1-k_2\right)^2\\
\langle\sigma\upsilon\rangle^{\gamma\gamma}_{C5,6}&=& \frac{4m_{\chi}^4}{\pi\Lambda_{f5}^6}\left(k_1 c_w^2 + k_2 s_w^2 \right)^2\\
\langle\sigma\upsilon\rangle^{\gamma Z}_{C5,6}&=& \frac{ 3(4m_{\chi}^2-m_{Z}^2)^3 c_w^2 s_w^2 }{ 32\pi m_{\chi}^2 \Lambda_{f5}^6 }\left(k_1-k_2\right)^2\\
\langle\sigma\upsilon\rangle^{\gamma\gamma}_{C7,8}&=& \frac{8m_{\chi}^4}{\pi\Lambda_{f5}^6}\left(k_1 c_w^2 + k_2 s_w^2 \right)^2\\
\langle\sigma\upsilon\rangle^{\gamma Z}_{C7,8}&=& \frac{ (4m_{\chi}^2-m_{Z}^2)^3 c_w^2 s_w^2 }{ 4\pi m_{\chi}^2 \Lambda_{f5}^6 }\left(k_1-k_2\right)^2.
\label{sigvs}
\end{eqnarray}
Numerical annihilation rates for our operators are sketched in Figure \ref{anns}.

\subsubsection{Collider Bounds and the FermiLAT line}

Assuming that the observed feature at $E_{\gamma}\approx 130\gev$ in
the FermiLAT photon spectrum is a monochromatic gamma-ray line due to
dark-matter annihiliation, the measurement of the annihilation
cross-section selects a region of parameter space
of our operators and allows for a specific prediction for a collider signal.

\begin{figure}[hbtp]
    \centering
    \includegraphics[width=1.0\linewidth]{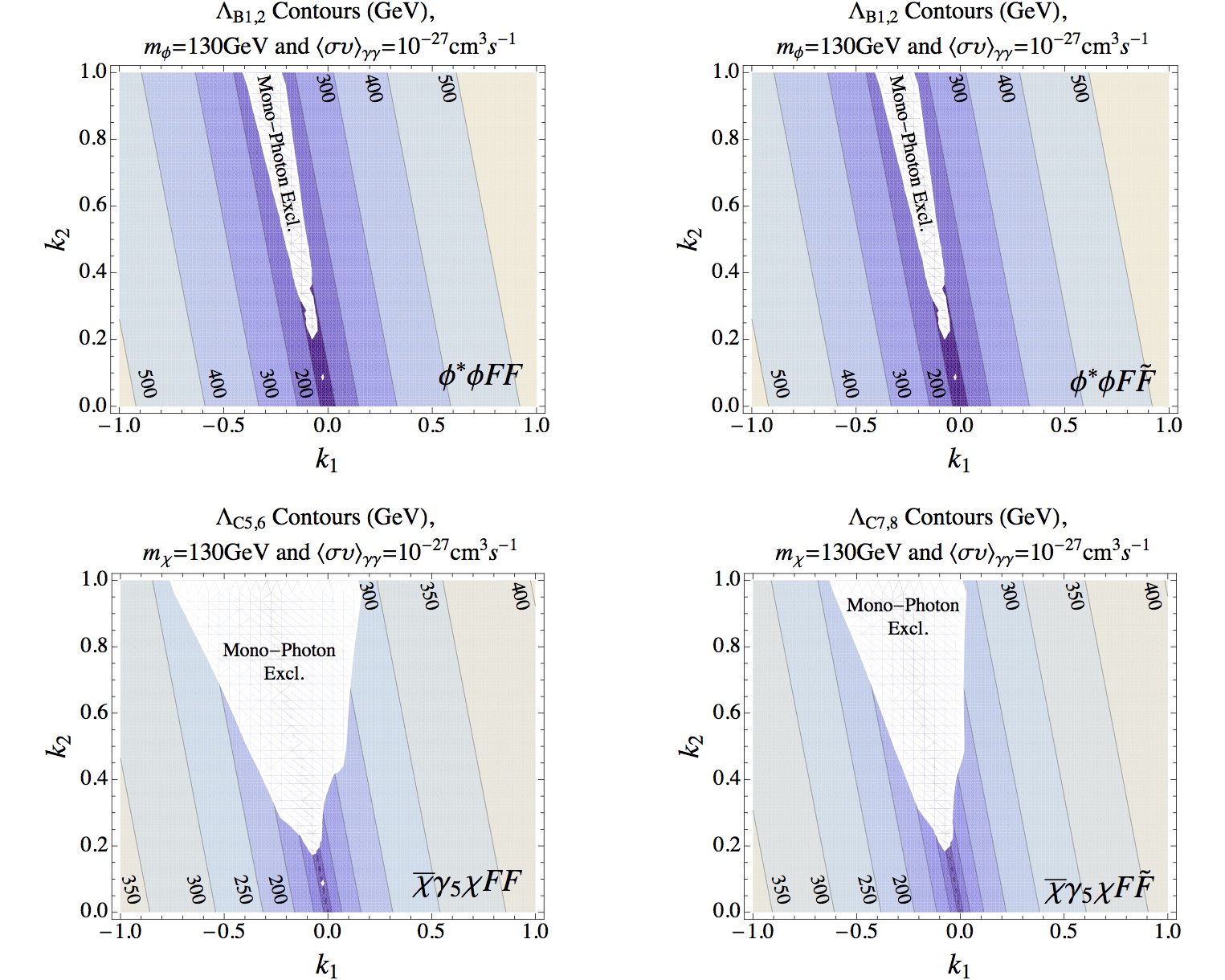}
    \caption{Contours of $\Lambda$ necessary for $m_{DM}=130\gev$ and $\sigv_{\gamma \gamma}=10^{-27}\rm{cm^3s^{-1}}$ for 
      various operators in the $k_2$ vs.\ $k_1$ plane. Unshaded regions are excluded by our monophoton analysis.}
    \label{exclgg}
  \end{figure}

\begin{figure}[hbtp]
    \centering
    \includegraphics[width=1.0\linewidth]{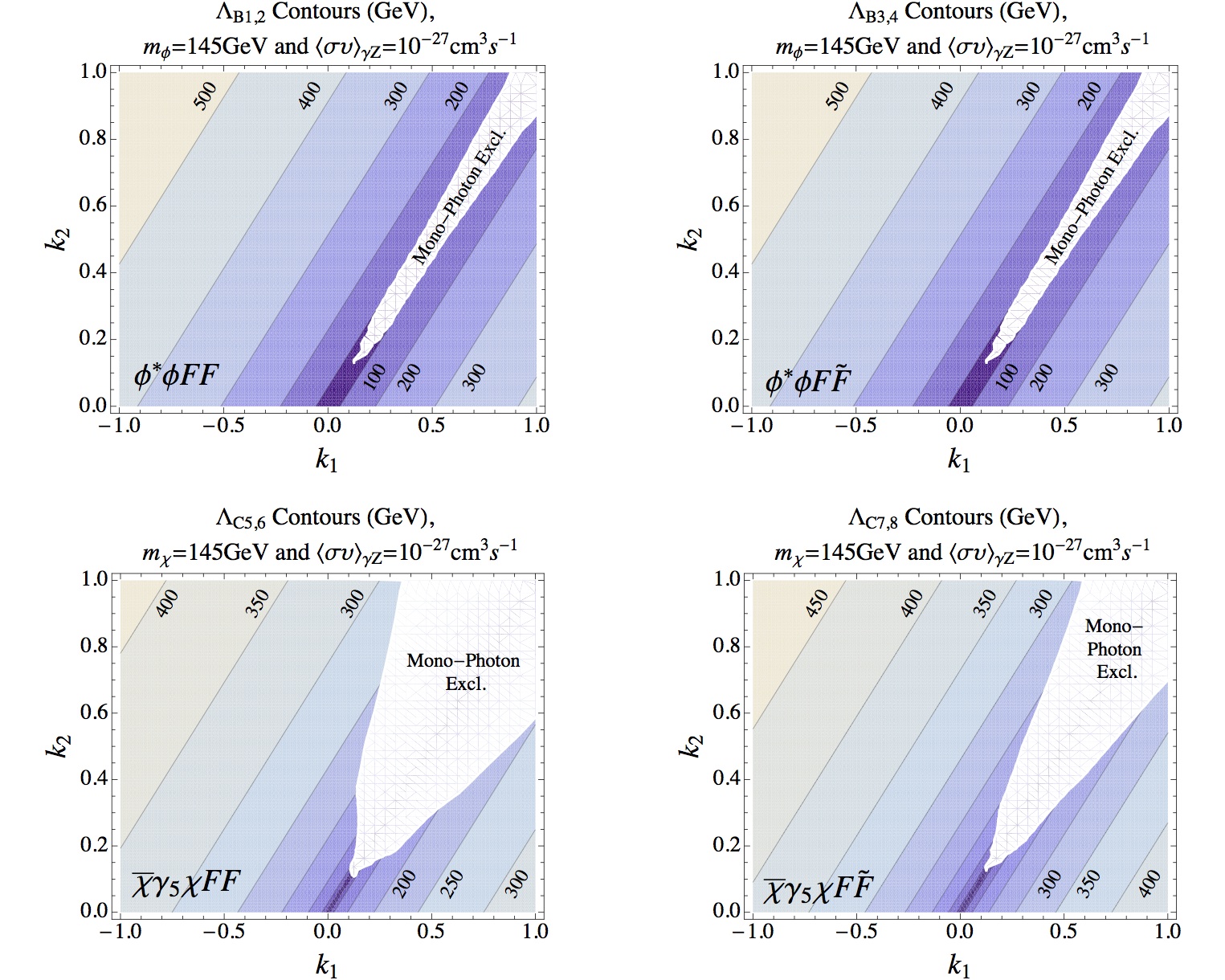}
    \caption{As in Figure \ref{exclgg} except for a scenario with $m_{DM}=145\gev$ and $\sigv_{\gamma Z}=10^{-27}\rm{cm^3s^{-1}}$.}
    \label{exclgz}
  \end{figure}

We first determine the regions of parameter space that give the line signal under two different hypotheses: $m_{DM}=130\gev$ with
$\sigv_{\gamma\gamma}=10^{-27}\rm{cm^3s^{-1}}$, and $m_{DM}=145\gev$ with $\sigv_{\gamma Z}=10^{-27}\rm{cm^3s^{-1}}$. This gives us a 
surface $\Lambda=\Lambda_{\textrm{line}}(k_i)$ which generates one of
these lines and can be immediately compared to the collider excluded region
$\Lambda\leq\Lambda_{\textrm{excl}}(k_i)$. The resulting allowed regions are shown in Figures \ref{exclgg}-\ref{exclgz}.

For both classes of DM we find that the desired line cross-section can be obtained for approximately electroweak-scale values of 
$\Lambda$ over most of the $k_1$ vs.\ $k_2$ plane. Since the cross-sections leading to $\gamma\gamma$ and $\gamma Z$ have different
dependences on our $k_1$ and $k_2$ parameters (Eqn. \ref{sigvs}), our $\Lambda$ contours are arranged quite differently for
the different final states in this plane. The $\Lambda$ values required to make an observable line drop sharply in regions where
the underlying $B$ and $W^0$ amplitudes interfere: $k_1=-t_w^2 k_2$ for $\gamma\gamma$ and $k_1=k_2$ for $\gamma Z$. 

We observe, as expected, that our mono-photon bounds rule out the bulk of these interference regions in our parameter space, leaving only 
the parameter space in the limit when \emph{both} $k_{1,2}\approx
0$. As noted above, the observed limits on operators with fermionic DM are relatively strong when compared with those on operators with
scalar DM. In approximate language, the bounds on scalar operators reach $\Lambda\lsim 150\gev$ (a number that is small compared to the
electroweak vev) while bounds on fermionic operators reach $\Lambda$ values at the several hundreds of $\gev$ levels (of the order of
the electroweak vev). This explains our observation in Figs.\ \ref{exclgg}-\ref{exclgz} that the collider bounds on our scalar operators
exclude essentially only regions of parameter space where some amount of tuning of $k_1$ and $k_2$ happens to reduce $\Lambda$ much below 
the weak scale ($\ie$, the interference regions), while collider bounds on the fermionic operators reach more general parts of our parameter
space.
 

\subsection{Conclusions}

In this work we have derived constraints on dark matter interactions
with photons in the context of a simply parameterized effective theory
framework. $\gamma+\missET$ bounds derived by the ATLAS collaboration for
dark matter interactions with quarks were recast to find bounds on our model 
for both scalar and fermionic dark matter scenarios. The kinematic differences in the two classes
of DM give bounds on the dimensionful scale of the effective operators that is more tightly constraining 
for fermionic DM than for the scalar case, roughly $\Lambda\gsim 300\gev$ and $\Lambda\gsim 150\gev$,
respectively.

We have also investigated the interplay between the collider data
and that from indirect detection searches for the energetic products
of dark matter annihilation in the galaxy. A putative FermiLAT signal
of DM annihilations to monochromatic gamma-rays results from our operator
setup on a particular subspace of its parameters. We have described 
collider constraints on this subspace, finding that, although much of 
parameter space can be excluded by the collider search, most of the 
space remains viable.

\subsection{Acknowledgements}

We acknowledge useful conversations with Tim Tait.
DW and AN are supported by grants from the Department of Energy
Office of Science. RC greatfully acknowledges the hospitality of
the KITP and was supported in part by the National Science Foundation
under PHY-0970173 and PHY11-25915.

\clearpage
\appendix


\begin{thebibliography}{99}

\bibitem{dmReview}
  G.~Bertone, D.~Hooper, J.~Silk,
  %``Particle Dark Matter: Evidence, Candidates and Constraints,''
  Phys.\ Rept.\ {\bf 405}, 279 (2005)
  [arXiv:0404175 [hep-ph]].
  %%CITATION = ARXIV:0404175;%%


\bibitem{fermiLatExcess}
  C.~Weniger,
  %``Tentative observation of a gamma ray line at the Fermi LAT,''
  AIP Conf.\ Proc.\ {\bf 1505}, 470 (2012)
  [arXiv:1210.3013 [astro-ph.HE]].
  %%CITATION = ARXIV:1210.3013;%%

\bibitem{fermisyst1}
  D.~Whiteson,
  %``Disentangling Instrumental Features of the 130 GeV Fermi Line,''
  JCAP {\bf 1211}, 008 (2012)
  [arXiv:1208.3677 [astro-ph.HE]].

\bibitem{fermisyst2}
 D.~P.~Finkbeiner, M.~Su and C.~Weniger,
  %``Is the 130 GeV Line Real? A Search for Systematics in the Fermi-LAT Data,''
  JCAP {\bf 1301}, 029 (2013)
  [arXiv:1209.4562 [astro-ph.HE]].

\bibitem{fermisyst3}
 D.~Whiteson,
  %``Searching for Spurious Solar and Sky Lines in the Fermi-LAT Spectrum,''
  arXiv:1302.0427 [astro-ph.HE].


\bibitem{atlasjet} 
  G.~Aad {\it et al.}  [ATLAS Collaboration],
  %``Search for dark matter candidates and large extra dimensions in events with a jet and missing transverse momentum with the ATLAS detector,''
  JHEP {\bf 1304}, 075 (2013)
  [arXiv:1210.4491 [hep-ex]].
  %%CITATION = ARXIV:1210.4491;%%


%\cite{Chatrchyan:2012me}
\bibitem{cmsjet}
  S.~Chatrchyan {\it et al.}  [CMS Collaboration],
  %``Search for dark matter and large extra dimensions in monojet events in pp collisions at sqrt(s)= 7 TeV,''
  JHEP {\bf 1209}, 094 (2012)
  [arXiv:1206.5663 [hep-ex]].
  %%CITATION = ARXIV:1206.5663;%%

\bibitem{atlasphoton}
  G.~Aad {\it et al.}  [ATLAS Collaboration],
  %``Search for dark matter candidates and large extra dimensions in events with a photon and missing transverse momentum in $pp$ collision data at $\sqrt{s}=7$ TeV with the ATLAS detector,''
  arXiv:1209.4625 [hep-ex].
  %%CITATION = ARXIV:1209.4625;%%


%\cite{Chatrchyan:2012tea}
\bibitem{cmsphoton}
  S.~Chatrchyan {\it et al.}  [CMS Collaboration],
  %``Search for Dark Matter and Large Extra Dimensions in pp Collisions Yielding a Photon and Missing Transverse Energy,''
  Phys.\ Rev.\ Lett.\  {\bf 108}, 261803 (2012)
  [arXiv:1204.0821 [hep-ex]].
  %%CITATION = ARXIV:1204.0821;%%


\bibitem{monoz} 
L.~M.~Carpenter, A.~Nelson, C.~Shimmin, T.~M.~P.~Tait and D.~Whiteson,
  %``Collider searches for dark matter in events with a Z boson and missing energy,''
  arXiv:1212.3352 [hep-ex].


\bibitem{Bell:2012rg} 
  N.~F.~Bell, J.~B.~Dent, A.~J.~Galea, T.~D.~Jacques, L.~M.~Krauss and T.~J.~Weiler,
  %``Searching for Dark Matter at the LHC with a Mono-Z,''
  Phys.\ Rev.\ D {\bf 86}, 096011 (2012)
  [arXiv:1209.0231 [hep-ph]].

\bibitem{Goodman:2010ku} 
  J.~Goodman, M.~Ibe, A.~Rajaraman, W.~Shepherd, T.~M.~P.~Tait and H.~-B.~Yu,
  %``Constraints on Dark Matter from Colliders,''
  Phys.\ Rev.\ D {\bf 82}, 116010 (2010)
  [arXiv:1008.1783 [hep-ph]].
  %%CITATION = ARXIV:1008.1783;%%


%\cite{Beltran:2010ww}
\bibitem{Beltran:2010ww} 
  M.~Beltran, D.~Hooper, E.~W.~Kolb, Z.~A.~C.~Krusberg and T.~M.~P.~Tait,
  %``Maverick dark matter at colliders,''
  JHEP {\bf 1009}, 037 (2010)
  [arXiv:1002.4137 [hep-ph]].
  %%CITATION = ARXIV:1002.4137;%%

\bibitem{Fox:2011pm} 
  P.~J.~Fox, R.~Harnik, J.~Kopp and Y.~Tsai,
  %``Missing Energy Signatures of Dark Matter at the LHC,''
  Phys.\ Rev.\ D {\bf 85}, 056011 (2012)
  [arXiv:1109.4398 [hep-ph]].
  %%CITATION = ARXIV:1109.4398;%%


%\cite{Cotta:2012nj}
\bibitem{Cotta:2012nj} 
  R.~C.~Cotta, J.~L.~Hewett, M.~P.~Le and T.~G.~Rizzo,
  %``Bounds on Dark Matter Interactions with Electroweak Gauge Bosons,''
  arXiv:1210.0525 [hep-ph].
  %%CITATION = ARXIV:1210.0525;%%

%\cite{Rajaraman:2012fu}
\bibitem{Rajaraman:2012fu} 
  A.~Rajaraman, T.~M.~P.~Tait and A.~M.~Wijangco,
  %``Effective Theories of Gamma-ray Lines from Dark Matter Annihilation,''
  arXiv:1211.7061 [hep-ph].
  %%CITATION = ARXIV:1211.7061;%%




%% Analysis

\bibitem{cls1} {A. Read},   J. Phys. G: Nucl. Part. Phys. {\bf 28}, 2693 (2002);
\bibitem{cls2} {T. Junk},  Nucl. Instrum. Methods A {\bf 434}, 425
  (1999).

%%% Simulation tools

\bibitem{madgraph}
  J.~Alwall, M.~Herquet, F.~Maltoni, O.~Mattelaer and T.~Stelzer,
  %``MadGraph 5 : Going Beyond,''
  JHEP {\bf 1106}, 128 (2011)
  [arXiv:1106.0522 [hep-ph]].
  %%CITATION = ARXIV:1106.0522;%%




\end{thebibliography}
\end{document}